\documentclass[aps,prb,twocolumn,superscriptaddress,longbibliography]{revtex4-2}

\usepackage{amsmath,amssymb,bm,mathtools,pifont}  
\usepackage{braket}                        
\usepackage{chemformula}                   
\usepackage{pifont}

\usepackage{graphicx}                      
\usepackage{overpic} 

\usepackage{tikz}                          
\usepackage{float}                         
\usepackage{svg}                           
\usepackage{epstopdf}                      

\usepackage{array}
\usepackage{multirow}
\usepackage{booktabs}                      
\usepackage{tabu}                          
\usepackage{array}

\usepackage[usenames,dvipsnames]{xcolor}   

\usepackage[	
citecolor=ForestGreen,
linkcolor=violet,
urlcolor=black,
]{hyperref} 
\usepackage{natbib}                        

\usepackage{latexsym}
\usepackage{comment}

\newcommand\redsout{\bgroup\markoverwith{\textcolor{red}{\rule[0.5ex]{2pt}{0.4pt}}}\ULon}
\newcommand\bluesout{\bgroup\markoverwith{\textcolor{blue}{\rule[0.5ex]{2pt}{0.4pt}}}\ULon}

\newcommand{\SPhide}[1]{}
\newcommand{\JDhide}[1]{}
\newcommand{\SKuhide}[1]{}

\begin{document}

\title{Critical SO(5) scaling of entanglement entropy at honeycomb lattice \\deconfined criticality}

\author{Sankalp Kumar}
\email{sankalpnimo@gmail.com}
\affiliation{Department of Physics, Indian Institute of Technology Bombay, Mumbai, MH 400076, India}

\author{Jonathan D'Emidio}
\email{jdemidio@utk.edu}
\affiliation{Department of Physics and Astronomy, University of Tennessee, Knoxville, TN 37996, USA}

\author{Sumiran Pujari}
\email{sumiran.pujari@iitb.ac.in}
\affiliation{Department of Physics, Indian Institute of Technology Bombay, Mumbai, MH 400076, India}

\begin{abstract}
The deconfined quantum critical point (DQCP) in square lattice $S=\frac{1}{2}$ quantum antiferromagnets has been extensively studied with a large body of evidence pointing to a weakly first-order transition scenario. 
Recent studies, which focused on entanglement at this nearly continuous DQCP in square lattice $J$-$Q$ models, have observed conflicting bipartite entanglement entropy scaling behavior depending on the details of the entanglement cut. 
One choice of the cut gave scaling coefficients in remarkable agreement with predictions from the unitary CFT corresponding to the putative DQCP. 
While another equally natural choice for the cut gave scaling coefficients in complete violation of unitary CFT that may be attributed to lack of scale invariance due to the known weakly first-order behavior of the model.
This motivates the exploration of DQCP behavior via entanglement measures in lattice models with distinct crystalline symmetries. 
Here we study a sign-free $S=\frac{1}{2}$ honeycomb lattice model that hosts a nearly continuous transition between antiferromagnetic N\'eel and three-fold valence-bond-solid ground states relevant to probing DQCP. 
Using large-scale quantum Monte Carlo simulations, we compute the R\'enyi entanglement entropy for a variety of bipartitions and test the CFT based description of the DQCP on the honeycomb lattice. 
For smooth bipartitions, we find no evidence of logarithmic corrections, in accordance with CFT, thereby essentially ruling out contributions from Goldstone modes. 
For subsystems with corners, CFT predicts universal logarithmic contributions, which we extract for corners with $60^\circ$ and $120^\circ$ angles and find close agreement with an emergent $\mathrm{SO}(5)$ CFT. 
While we observe scaling consistent with a critical system in the majority of cases, we also demonstrate an intriguing counterexample of the hexagon subsystem that exhibits a subtle period three oscillation. 
This results in three separate finite-size series, where the sign of the logarithmic term apparently changes depending on the series.
\end{abstract}

\maketitle


\section{Introduction}
\label{sec:intro}
Deconfined quantum criticality represents an important class of quantum phase transitions that lie beyond the conventional Landau-Ginzburg-Wilson (LGW) paradigm. 
Within the LGW framework, continuous phase transitions involving spontaneous symmetry breaking are characterized by an order parameter associated with the broken symmetry. 
When two phases break unrelated symmetries, a direct continuous transition between them is generically forbidden. 
One rather expects either a first-order transition or the appearance of an intermediate phase. 
Deconfined quantum criticality (DQCP) however provides a mechanism for a direct continuous transition between phases with distinct broken symmetries. 
This idea or framework was originally proposed by Senthil \textit{et al.}~\cite{Senthil_etal_science_2004} in the setting of $S=\frac{1}{2}$ quantum antiferromagnets which will also be the setting in this study.
At such a critical point, the theory predicts the emergence of fractionalized degrees of freedom and emergent gauge fields, together with enhanced symmetries that are absent in the microscopic model. 
Over the past two decades, DQCPs have thus been the subject of extensive theoretical and numerical investigation.

A continuous transition between a N\'eel antiferromagnet and a valence-bond-solid (VBS) phase in two-dimensional quantum spin-$\frac{1}{2}$ systems has become the paradigmatic example of DQCP. 
In brief, the VBS phase may be viewed as a confined phase of emergent spin-$\frac{1}{2}$ excitations, known as spinons, which bind into singlet pairs~\cite{Levin_Senthil_2004}. 
In the VBS phase these singlets order and break the lattice symmetries. 
Upon melting of the VBS order, the singlet pairs unbind leading to the deconfinement of spinons precisely at the DQCP point. 
Upon tuning further away from this VBS melting point, the spinons immediately condense to give rise to the N\'eel ordered phase.
Ref.~\cite{Levin_Senthil_2004}'s insight was in showing that the defects of the VBS phase indeed are spin-$\frac{1}{2}$ degrees of freedom located at the core of the VBS domain walls, i.e. they possess the quantum numbers of the emergent spinon which is a pre-condition for the above picture to be viable.

We note here that the original proposal~\cite{Senthil_etal_science_2004,Senthil_etal_PRB_2004} was formulated in an alternate or dual description starting from the N\'eel ordered side and its (topological) defect structure in terms of a noncompact $\mathrm{CP}^{N-1}$ field theory with an emergent $U(1)$ gauge structure.
There exists yet another formulation of DQCP as a non-linear sigma model in terms of a combined 5-component order parameter that is made up of both the N\'eel and VBS order parameters proposed in Ref.~\cite{Senthil_Fisher_2006,Tanaka_Hu_2005}.
This automatically implies an emergent SO(5) symmetry at the DQCP which leads to the expectation of an associated SO(5) CFT description.
This will form the crux of the investigations in this study.
It will also be useful to keep in mind the VBS melting picture of Ref.~\cite{Levin_Senthil_2004}  based formulation of the DQCP for some of the discussions later in the paper.

DQCP on general grounds is to be expected in the presence of frustration either through lattice geometry or through competing interactions that disfavor N\'eel order.
For large-scale studies through quantum Monte Carlo simulations, geometrically frustrated lattice models suffer from the infamous sign problem. 
Thus models with competing interactions that do not create sign problems have become the playground of choice and go by the name of designer models.
The first sign-problem-free lattice realization of this physics was provided by the square-lattice spin-$\frac{1}{2}$ ``$J$-$Q$" model introduced by Sandvik~\cite{Sandvik_2007}. 
It has nearest-neighbor Heisenberg interactions $J\sum_{\langle ij\rangle}\mathbf{S}_i\cdot\mathbf{S}_j$ favoring N\'eel order which competes with four-spin interactions of the form $Q_2 \sum_{\langle ijkl\rangle}(\mathbf{S}_i\cdot\mathbf{S}_j)(\mathbf{S}_k\cdot\mathbf{S}_l)$ on square plaquettes which favor VBS order. 
The subscript in $Q_2$ indicates that the designer term is formed from products of two Heisenberg exchange terms.
This model and the associated $J$-$Q$ model class has enabled many large-scale quantum Monte Carlo (QMC) studies of DQCP to date. 

Numerical work on the nature of the transition in these models has produced distinct conclusions, resulting in a surprisingly rich story.  
Most of the literature has focused on the (nearly) continuous scenario for the N\'eel-VBS transition \cite{Sandvik_2007,Lou2009:AFtoVBSsun,Sandvik2010:ContinuousJQlog,Pujari_Damle_Alet_2013,Shao_Guo_Sandvik_Science_2016,Sandvik2020:Consistent,Zhao2020:Helical,Harada2013:DQCPsmallN,Block2013:Fate}, however the best current evidence convincingly shows rather a very weakly first-order transition \cite{Kuklov2008:DCPfirstorder,Jiang2008:FirstOrder,Chen2013:DCPflow,demidio2021diagnosing,Takahashi2024:multicritical}. 
The question has now evolved into why there is such an anomalously weak discontinuity, which can either be due to the presence of a nearby complex fixed point (occuring at complex parameter values)\cite{Wang2017:DQCPsymm,Gorbenko2018:Walking,Gorbenko2018:Walking2,Ma2020:DQCPpseudo,Nahum2020:Note,Zhou_etal_PRX_2024}, as is known to happen in the $Q$-state Potts model \cite{Gorbenko2018:Walking,Gorbenko2018:Walking2} and SU($N$) spin chains \cite{Kumar2026:PseudoSUN}, or a real fixed point in a region that is inaccessible due to the sign problem \cite{Takahashi2024:multicritical}. 
Despite the weakly first-order nature of the transition in the model, it is still possible to ask meaningful questions about the putative nearby critical point due to its extremely close proximity.   

This has led to studying DQCP through the lens of entanglement entropy (EE) in order to see if universal features of the nearby critical point are reflected in this quantity as well. 
At criticality, where scale invariance emerges, the low-energy physics is described by conformal field theory (CFT), which predicts universal scaling forms for the EE. 
In particular, the EE involving different bipartitions of the system encodes universal information in subleading corrections beyond the leading area-law contribution. 
CFT predicts that, in two-dimensional quantum systems at a second-order critical point, the EE of a subsystem with boundary length $l$ scales as~\cite{Fradkin_etal_2006,Casini_Huerta_2007}
\begin{equation}
S(l) = a\, l - b \ln l + c ,
\label{eq:2dCFT_form}
\end{equation}
where $a$ and $c$ are nonuniversal constants, while $b$ is a universal coefficient characterizing the underlying CFT. 
For subsystems with sharp corners, the coefficient $b$ receives additive universal contributions, $b = \sum_i b_{\theta_i} \,$, where $\theta_i$ denote the corner opening angles of the subsystem. 
In a pure state, since the EE of the subsystem and its complement are equal by definition, we can conclude that $b_{2\pi-\theta} = b_{\theta}$. 
The opening angle is thus conventionally chosen as $0<\theta_i<\pi$. 
For smooth subsystems without corners, the universal logarithmic contribution is predicted to vanish (equivalently $b_\pi=0$). 
In the presence of corners, the corner coefficients $b_\theta$ are predicted to be universal and satisfy the inequality $b_\theta > b_\pi = 0$.
The positivity of $b_\theta$ and its vanishing for smooth boundaries therefore provide a manifestation of the nearby CFT, despite the very small first-order discontinuity present in lattice models of DQCP.

EE-based studies have revealed several intriguing aspects of the DQCP. In a study of a closely related square-lattice $J$-$Q_3$ model~\cite{Zhao_etal_2022}, the extracted corner coefficients were found to violate the expected positivity constraints, and a nonzero logarithmic term was observed even for smooth subsystems \cite{Song2024:subleadingEE} whereas CFT predicts that it should vanish. 
These results were interpreted as evidence against a conventional unitary CFT description of the DQCP. 
Other studies of the DQCP furthermore showed consistency with the presence of SO(5) goldstone modes in EE~\cite{Deng2024:Diagnosing} and entanglement spectrum~\cite{Mao2026:Detecting} data, indicating a weak ordered moment at the transition. 
However, another study demonstrated that the entanglement entropy depends on the microscopic geometry of the bipartition and that the apparent violations arose from the use of ``improper" subsystem geometries. 
When subsystems were chosen according to the ``proper bipartitioning" heuristic~\cite{Jon_sandvik_Properbipartition_2024}, the corner coefficients satisfied the expected CFT constraints and were found to be consistent with an emergent $\mathrm{SO}(5)$-symmetric CFT~\cite{Whitsitt_etal_2017,Helmes_etal_2016}.

The idea behind proper bipartitioning is that the subsystem boundary should treat all degenerate VBS ground states on an equal footing.
On the square lattice, this corresponds to choosing a boundary that cuts the same number of singlets in each of the four degenerate columnar VBS configurations.
Such bipartitions were described as ``tilted" bipartitions to distinguish from the more conventional ``straight" bipartitions.
For illustrative figures showing this, see Fig. (1) of Ref.~\cite{Jon_sandvik_Properbipartition_2024}.
While this observation suggests that microscopic details of the bipartition can influence entanglement diagnostics even at criticality, the rationale remains somewhat ad hoc. 
The VBS order parameter after all vanishes at the critical point through the proliferation of VBS domain wall defects.
It is therefore unclear why properties associated with ordered VBS configurations should control universal entanglement contributions.

An alternative explanation was put forward in Ref.~\cite{Zhu_surface_criticality}, where it was argued that the difference in EE scaling between straight and tilted lattice bipartitions originates from the types of edges generated by the bipartition. 
In particular, the tilted lattice geometry can host additional contributions to the bulk EE from gapless edge modes associated with ``extraordinary surface criticality'' at the DQCP. 
Taken together, these observations raise a broader question: to what extent are entanglement-based diagnostics at the DQCP governed by universal critical physics, and to what extent are they influenced by microscopic features of the bipartition geometry? 
A more complete theoretical understanding of the role of bipartition geometry therefore remains lacking.

To address this question, we wish to examine independent realizations of the N\'eel-VBS transition in alternative lattice geometries. 
A natural choice is the honeycomb lattice, which has been shown to host a N\'eel-VBS transition pertinent to DQCP physics~\cite{Pujari_Damle_Alet_2013}. 
While sharing many qualitative features with square-lattice models, the honeycomb system differs in its lattice geometry and topological defect structure, making it a valuable platform for probing both the universality of entanglement diagnostics and the proposed $\mathrm{SO}(5)$ CFT description of the DQCP.
By systematically analyzing a variety of bipartitions of the honeycomb lattice, we investigate whether the N\'eel-VBS transition in the honeycomb $J$-$Q$ model exhibits entanglement signatures consistent with a CFT description with emergent $\mathrm{SO}(5)$ symmetry. 
Our key findings are as follows: 
We find that the EE at this transition follows the scaling form in Eq.~\ref{eq:2dCFT_form} in all cases. 
For smooth bipartitions, no logarithmic corrections to the area law are found, in agreement with CFT.
For bipartitions with corners, such as triangular and hexagonal geometries, logarithmic contributions are found whose corner coefficients are in remarkable agreement with a large-$N$ prediction from a ``$\mathrm{SO}(5)$''-based CFT.

There is an important caveat for putting quotation marks around $\mathrm{SO}(5)$ above.
The predictions are actually taken from $\mathrm{O}(N)$ Wilson-Fisher theory at criticality in $2+1$ dimensions at leading order ($\propto N$), i.e. it is equal to the value obtained for an $N$-component free scalar theory at Gaussian level~\cite{Whitsitt_etal_2017,Helmes_etal_2016}. 
Note that the DQCP is rather described by a non-linear sigma model in terms of an $\mathrm{SO}(5)$ order parameter as mentioned earlier along with a topological Wess-Zumino-Witten term and \emph{not} by a $\mathrm{O}(N)$ Wilson-Fisher theory.
The argument is that the leading order values would be the same for \emph{both} the theories, and the differences would manifest as lower-order ``corrections''.
Since $N=5$, the expectation is that the leading order value proportional to $N$ would be a good ballpark estimate for the ``true'' CFT value and would serve as a good point of comparison for the numerically observed scaling coefficients.
We will thus refer to the theory values in the rest of the paper as ``5-component'' values keeping in mind this large-$N$ caveat.

Last but not the least, we find an interesting subtlety in the case of hexagonal subsystems, which show a period-3 oscillation with the system size. 
This results in three distinct series, where depending on the series, the logarithmic term may change sign.
This is an important result from this work that we believe represents a genuine puzzle for field theory.
The remainder of this article is organized as follows. 
Sec.~\ref{sec:model} introduces the spin-$\frac{1}{2}$ honeycomb $J$-$Q$ model that we will investigate in context of DQCP, and discuss its differences from the square-lattice $J$-$Q$ models. 
Sec.~\ref{sec:QMC_method} outlines the QMC method used to compute entanglement entropy in this work. 
Sec.~\ref{sec:results} presents the entanglement results for various bipartitions with different types of edges and analyze their consistency with an emergent $\mathrm{SO}(5)$ CFT.
Sec.~\ref{sec:more_discussion} cross-examines the proper bipartitioning and surface criticality notions in light of our results.
Finally, Sec.~\ref{sec:conclusion} and Table~\ref{tab:summary_b} therein summarizes all our findings.


\section{The honeycomb $J$-$Q$ Model}
\label{sec:model}

We study the following spin-$\frac{1}{2}$ ``$J$-$Q$" model on the honeycomb lattice with the Hamiltonian 
\begin{equation}
H = -J \sum_{\langle ij \rangle} P_{ij}
    - Q \sum_{\langle ijklmn \rangle}
    P_{ij} P_{kl} P_{mn} ,
    \label{eq:JQ3_model}
\end{equation}
expressed in terms of spin-$\frac{1}{2}$ singlet projectors $P_{ij} = \tfrac{1}{4} - \mathbf{S}_i \cdot \mathbf{S}_j$ on the (nearest-neighbor) bond $\langle ij \rangle$. 
Here $\mathbf{S}_i = (S^x_i,S^y_i,S^z_i)$ are the standard spin-$\frac{1}{2}$ operators or generators on site $i$ of the honeycomb lattice.
The first term corresponds to an antiferromagnetic Heisenberg exchange ($J>0$) favoring N\'eel order.
The second term is a six-spin interaction ($Q>0$) defined on elementary hexagonal plaquettes $\langle ijklmn \rangle$ that stabilizes valence-bond-solid (VBS) order. 
The bonds $\{\langle ij \rangle,\langle kl \rangle,\langle mn \rangle\}$ entering the second term are of nearest-neighbor type and do not share any common sites.
There are two such possibilities for each hexagonal plaquette; see Fig.~1 of Ref.~\cite{Pujari_Damle_Alet_2013} for reference.

This model was previously claimed to exhibit a direct continuous quantum phase transition between a N\'eel antiferromagnet and a VBS phase at a critical coupling $(Q/J)_c = 1.190(6)$~\cite{Pujari_Damle_Alet_2013} sharing several features with the square-lattice $J$-$Q$ model transitions. 
However, in light of the present consensus that the square lattice $J$-$Q$ transition is weakly first-order, we believe the honeycomb $J$-$Q$ model transition is also of a weakly first-order kind. 
Despite this, numerical studies indicate that these transitions are overall consistent with the scenario of a DQCP. 
The common features include large anomalous dimensions of the two (N\'eel and VBS) order parameters and (close to) $U(1)$-symmetric histograms of the two-component complex VBS order parameter, as well as logarithmic violations in the conventional finite-size scaling of the spin stiffness.
This is apart from the inference of a single QCP between the symmetry unrelated phases as seen through robust Binder and correlation ratio crossings.
It has furthermore been shown that the above scenario continues to hold in the presence of perturbations except for an expected change in the location of the QCP $(Q/J)_c$~\cite{Pujari_Damle_Alet_2015}. 

Despite these similarities, the honeycomb $J$-$Q$ model differs in important ways from its square-lattice counterpart. 
Most notably, the ``columnar" VBS phase on the honeycomb lattice is threefold degenerate, corresponding to Kekul\'e-like ordering patterns at wavevector $\left(2\pi/3,-2\pi/3\right)$, in contrast to the fourfold degenerate columnar VBS phase on the square lattice at $(\pi,0) \text{ or }(0,\pi)$.
From the field theory side, this results from the difference in the nature of the topological defects in the N\'eel order parameter field in the low-energy theory~\cite{Senthil_etal_PRB_2004,Sachdev_QPTbook_2011}.
For the honeycomb lattice, these topological defects are three-fold monopole operators that act as seeds of VBS order upon their condensation. 
Compared to the four-fold monopole operators of the square lattice, the three-fold operators carry a smaller scaling dimension~\cite{Murthy1990:Hedgehog,Block2013:Fate}, and are thus more relevant at tree level.

Nevertheless, Ref.~\cite{Pujari_Damle_Alet_2015} showed a near-marginal behavior for these topological defects near the critical point. 
This had manifested as a pronounced, numerically observable threefold anisotropy in the phase of the (complex) VBS order parameter.
This was in contrast to an analogous (fourfold) anisotropy in the square lattice $J$-$Q$ model which was barely observable in the numerical studies. 
Furthermore, the honeycomb $J$-$Q$ model's threefold anisotropy was shown to decay very slowly with system size, indicating that lattice anisotropies persist over a wide range of length scales.
This was interpreted as the requisite dangerous irrelevance of the tripled monopoles for a viable DQCP in Ref.~\cite{Pujari_Damle_Alet_2015}.
Here we reinterpret this state of affairs, based on the known weakly first-order nature for the square lattice case, as a more pronounced version of a weakly first-order transition in the honeycomb model~\cite{footnote_binder_dips}. 
This would be in line with the scaling dimension considerations pointed out above. 
Importantly, as we will show, the effect is still weak enough to draw meaningful critical properties from  entanglement entropy at the honeycomb transition.


\section{Quantum Monte Carlo method}
\label{sec:QMC_method}

Conventional quantum Monte Carlo (QMC) techniques do not provide direct access to the von Neumann entanglement entropy. 
Nevertheless, significant methodological advances have enabled the efficient computation of R\'enyi entanglement entropies within the QMC framework based on replica methods~\cite{Melko_Kallin_Hasting_2010,Hastings_Gonzalez_Kallin_Melko_2010,Humeniuk_Roscilde_2012,Inglis_Melko_2013,Luitz_Plat_Laflorencie_Alet_2014,Kulchytskyy_etal_2015,Alba_2017,Jon_PRL_2020}. 
In this work, we consider the second R\'enyi entanglement entropy associated with a subsystem $A$ of the full system $A \cup \bar{A}$, defined as
\begin{equation}
S_2 = -\ln \left( \mathrm{Tr}\,\rho_A^2 \right),
\end{equation}
where $\rho_A$ denotes the reduced density matrix of subsystem $A$, obtained by tracing out the degrees of freedom in its complement $\bar{A}$:
\begin{equation}
\rho_A = \frac{\mathrm{Tr}_{\bar{A}}\!\left(e^{-\beta H}\right)}{\mathrm{Tr}\!\left(e^{-\beta H}\right)},
\end{equation}
where $\beta$ is the inverse temperature $\beta =(k_B T)^{-1}$. $\beta \rightarrow \infty$ gives the reduced density matrix of $A$ in the ground state. 
We follow the standard practice of choosing $\beta$ sufficiently large such that contributions from excited states are sufficiently suppressed, ensuring that $\rho_A$ effectively corresponds to the reduced density matrix of the ground state. In this work, we set $\beta = L$ (and $J=1$).
In terms of replicas, the second R\'enyi entanglement entropy can be written as a ratio of replica partition functions \cite{Calabrese_Cardy_2004},
$e^{-S_2} = \frac{Z_A}{Z_{\varnothing}}$,
where
$Z_X \equiv \mathrm{Tr}_X\!\left[\left(\mathrm{Tr}_{\bar{X}} e^{-\beta H}\right)^2\right]$, with $X=\varnothing$ denoting the empty set and $X=A$ the subsystem of interest. 

State-of-the-art QMC approaches evaluate this ratio efficiently through the introduction of generalized replica partition functions that interpolate between $Z_{\varnothing}$ and $Z_A$ via a tuning parameter.
One such construction, introduced in \cite{Jon_PRL_2020}, is
\begin{equation}
\mathcal{Z}(\lambda) = \sum_{B \subseteq A} \lambda^{N_B} (1-\lambda)^{N_A-N_B} Z_B ,
\end{equation}
where $B \subseteq A$, and $N_A$ and $N_B$ denote the number of sites in $A$ and $B$, respectively.
The parameter $\lambda$ controls the probability that a site in $A$ is traced only once, yielding $\mathcal{Z}(0)=Z_{\varnothing}$ and $\mathcal{Z}(1)=Z_A$.

In order to compute $S_2$, we employ the equilibrium method introduced in Ref.~\cite{Jon_sandvik_Properbipartition_2024}. In this approach, the ratio $Z_A/Z_{\varnothing} = \mathcal{Z}(1)/\mathcal{Z}(0)$ is decomposed into a product of incremental ratios,
\begin{equation}
\frac{\mathcal{Z}(1)}{\mathcal{Z}(0)}
= \prod_{i=0}^{n} \frac{\mathcal{Z}(\lambda_{i+1})}{\mathcal{Z}(\lambda_i)} ,
\end{equation}
where $\lambda_0 = 0$ and $\lambda_{n+1} = 1$.
Each ratio $\mathcal{Z}(\lambda_{i+1})/\mathcal{Z}(\lambda_i)$ is computed independently using a reweighting procedure,
\begin{equation}
\frac{\mathcal{Z}(\lambda_{i+1})}{\mathcal{Z}(\lambda_i)}
= \left\langle
\left( \frac{\lambda_{i+1}}{\lambda_i} \right)^{N_B}
\left( \frac{1-\lambda_{i+1}}{1-\lambda_i} \right)^{N_A-N_B}
\right\rangle_{\lambda_i},
\end{equation}
where the expectation value is taken with respect to the equilibrium ensemble defined by $\mathcal{Z}(\lambda_i)$.
We do the above in the stochastic series expansion (SSE) framework of QMC \cite{sandvik_book}.
For more details on the measurement of R\'enyi entropies in the SSE-QMC, see Ref.~\cite{Jon_PRL_2020,Jon_sandvik_Properbipartition_2024}, including the supplemental materials therein.


\begin{figure*}[t]
\centering
\begin{minipage}{0.48\linewidth}
\begin{overpic}[width=0.48\linewidth,clip=true,trim=0 0 0 2cm]{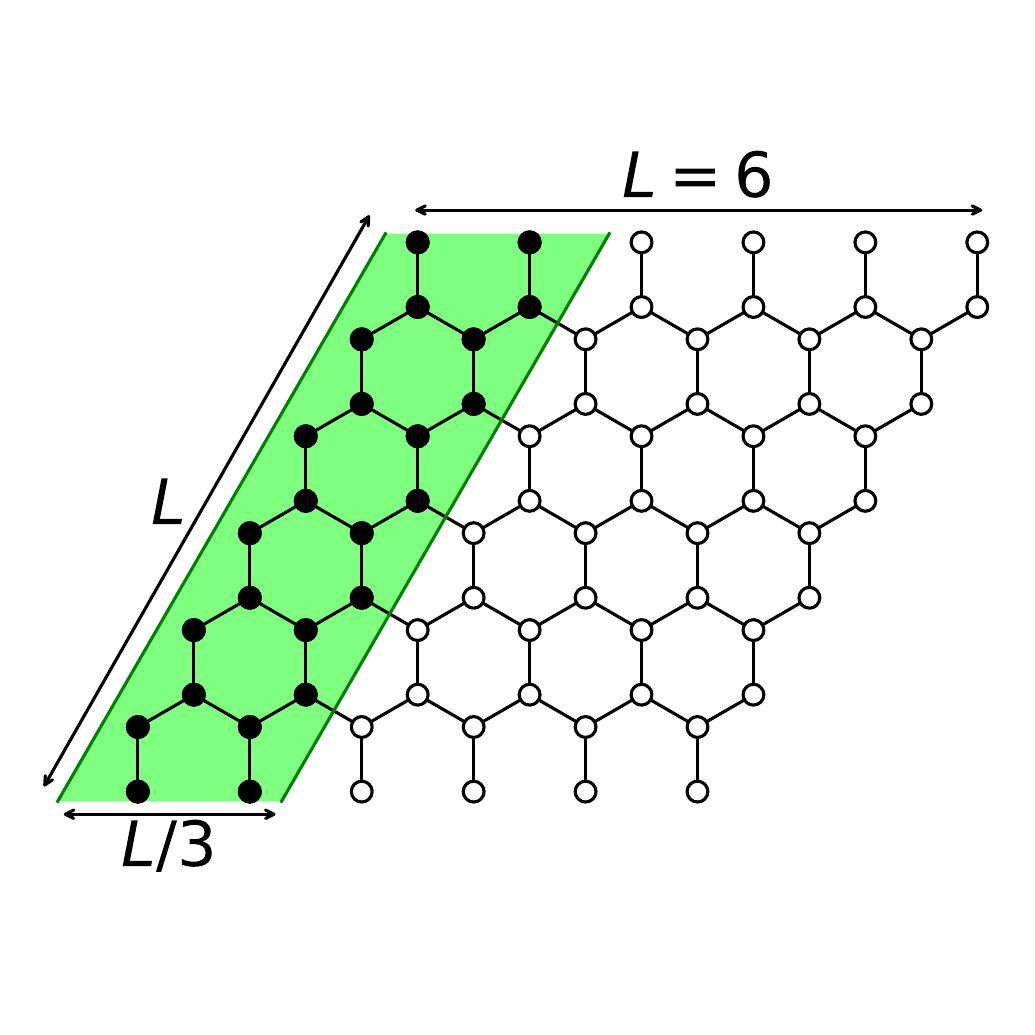}
\put(4,80){(a)}
\end{overpic}
\begin{overpic}[width=0.48\linewidth,clip=true,trim=0 0 0 2cm]{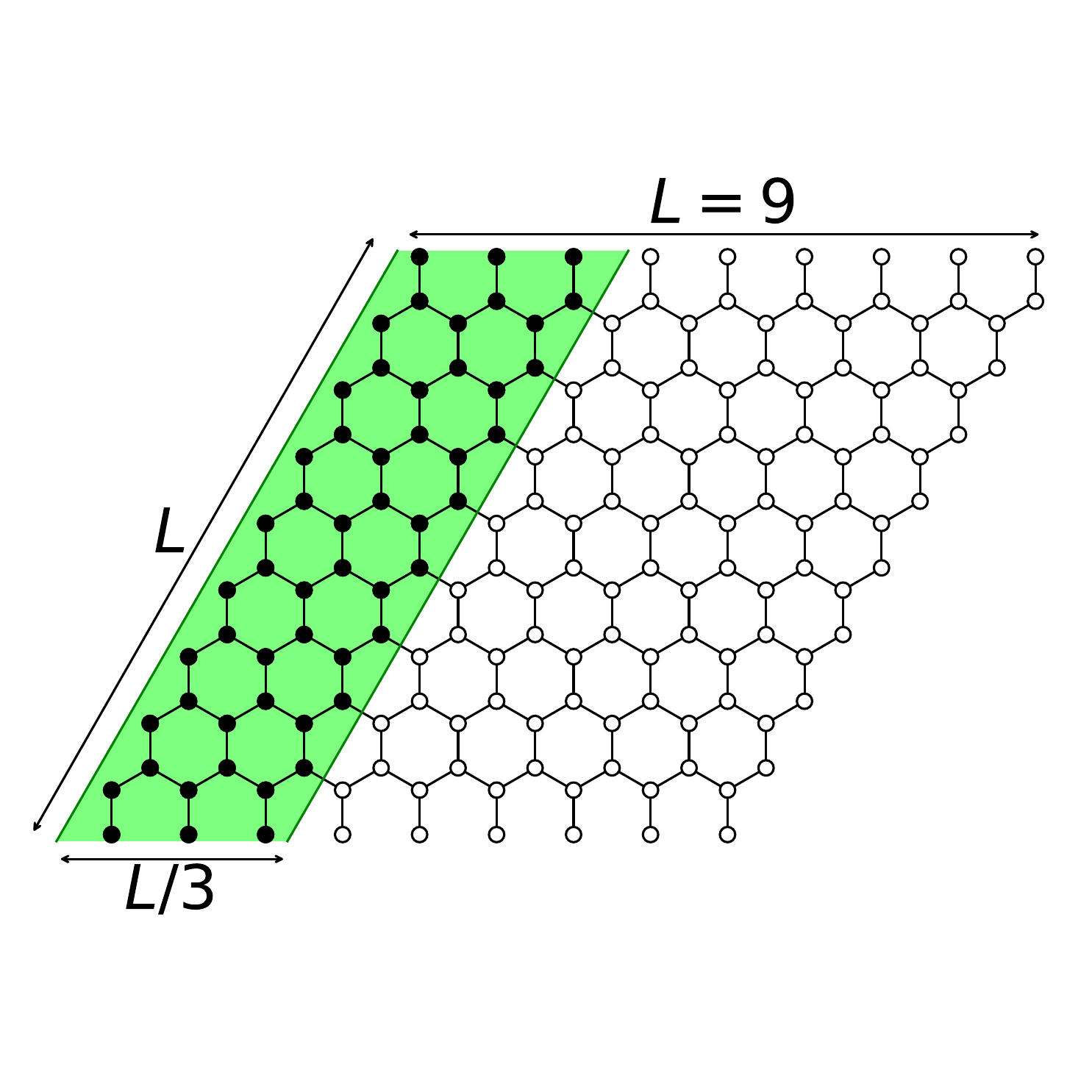}
\put(4,80){(b)}
\end{overpic}
\begin{overpic}[width=0.48\linewidth,clip=true,trim=0 0 0 2cm]{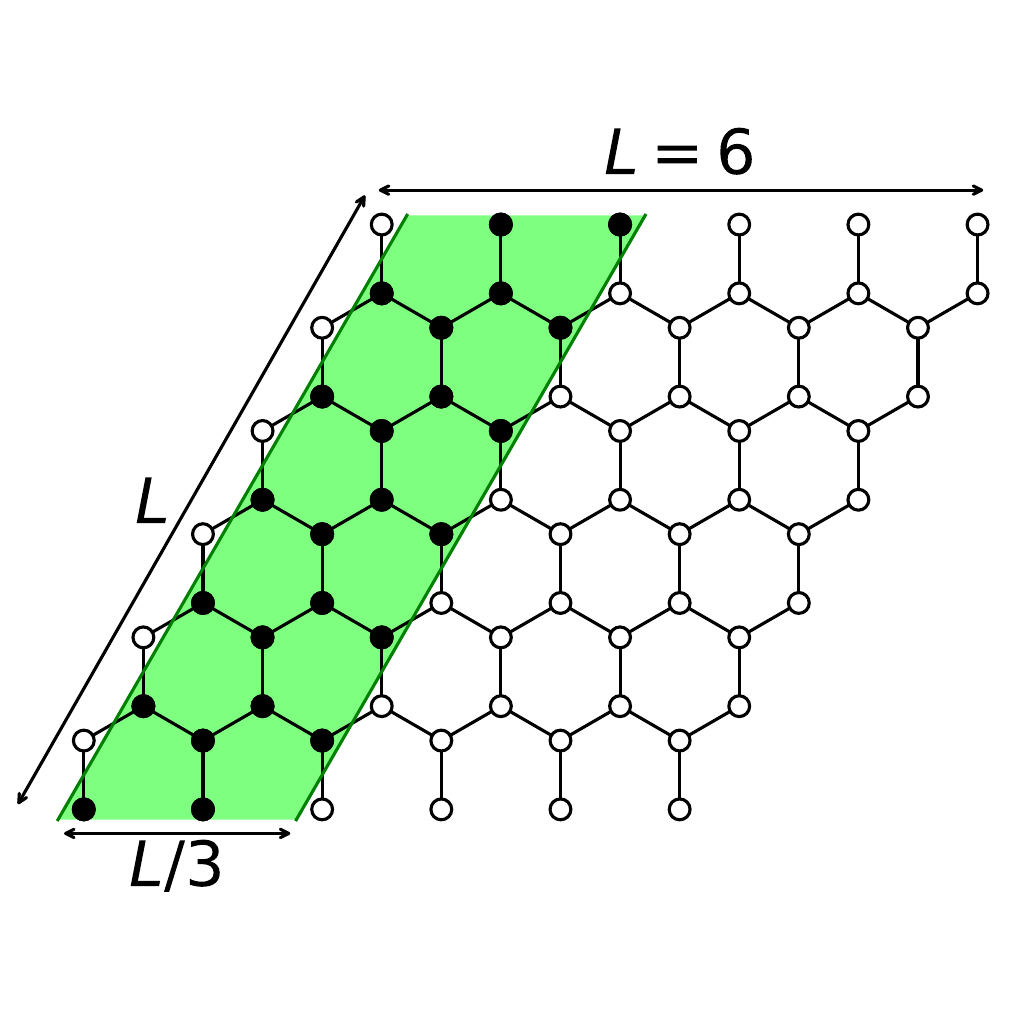}
\put(4,80){(c)}
\end{overpic}
\begin{overpic}[width=0.48\linewidth,clip=true,trim=0 0 0 2cm]{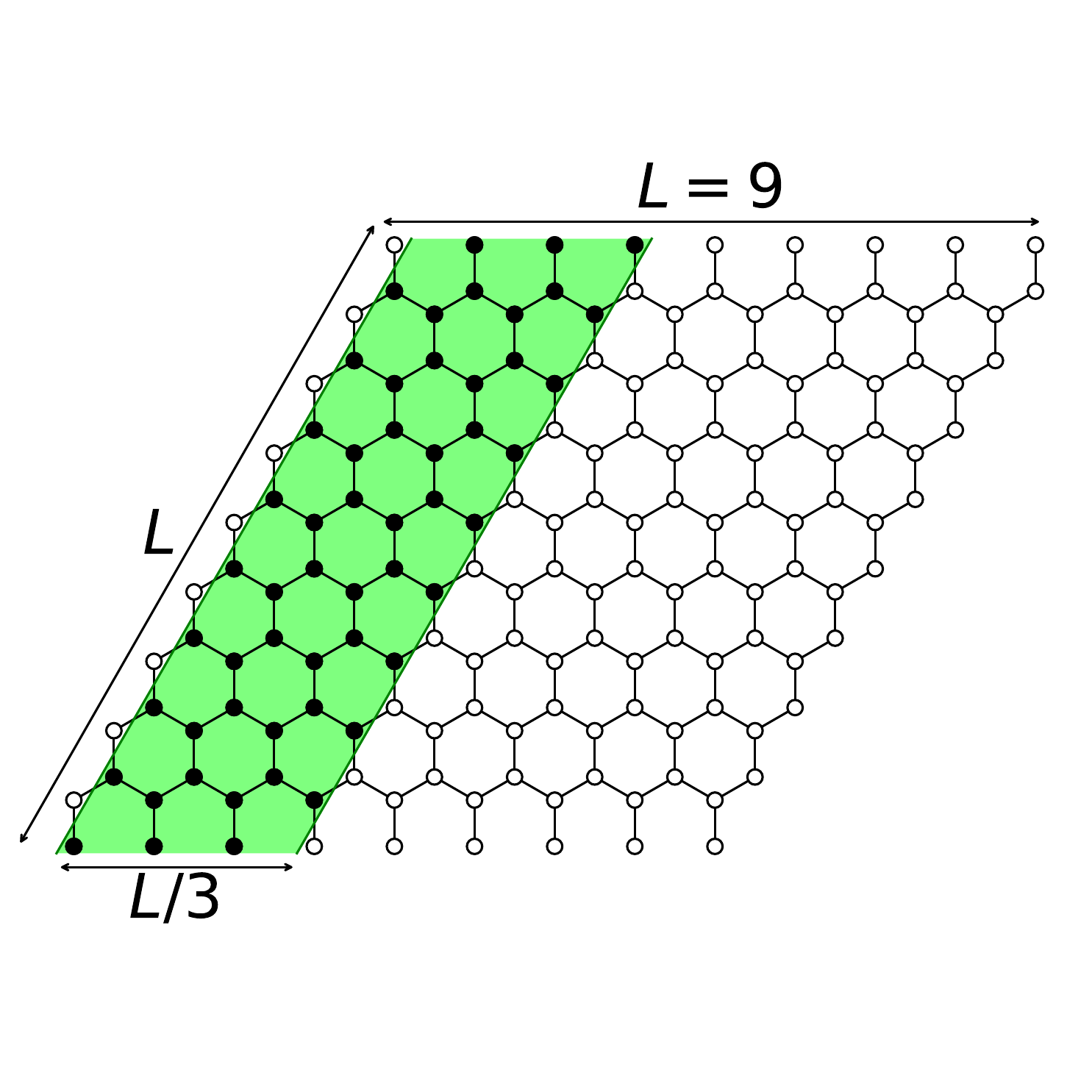}
\put(4,80){(d)}
\end{overpic}
\end{minipage}
\begin{minipage}{0.48\linewidth}
\begin{overpic}[width=0.8\linewidth,clip=true,trim=0 0 0 0cm]{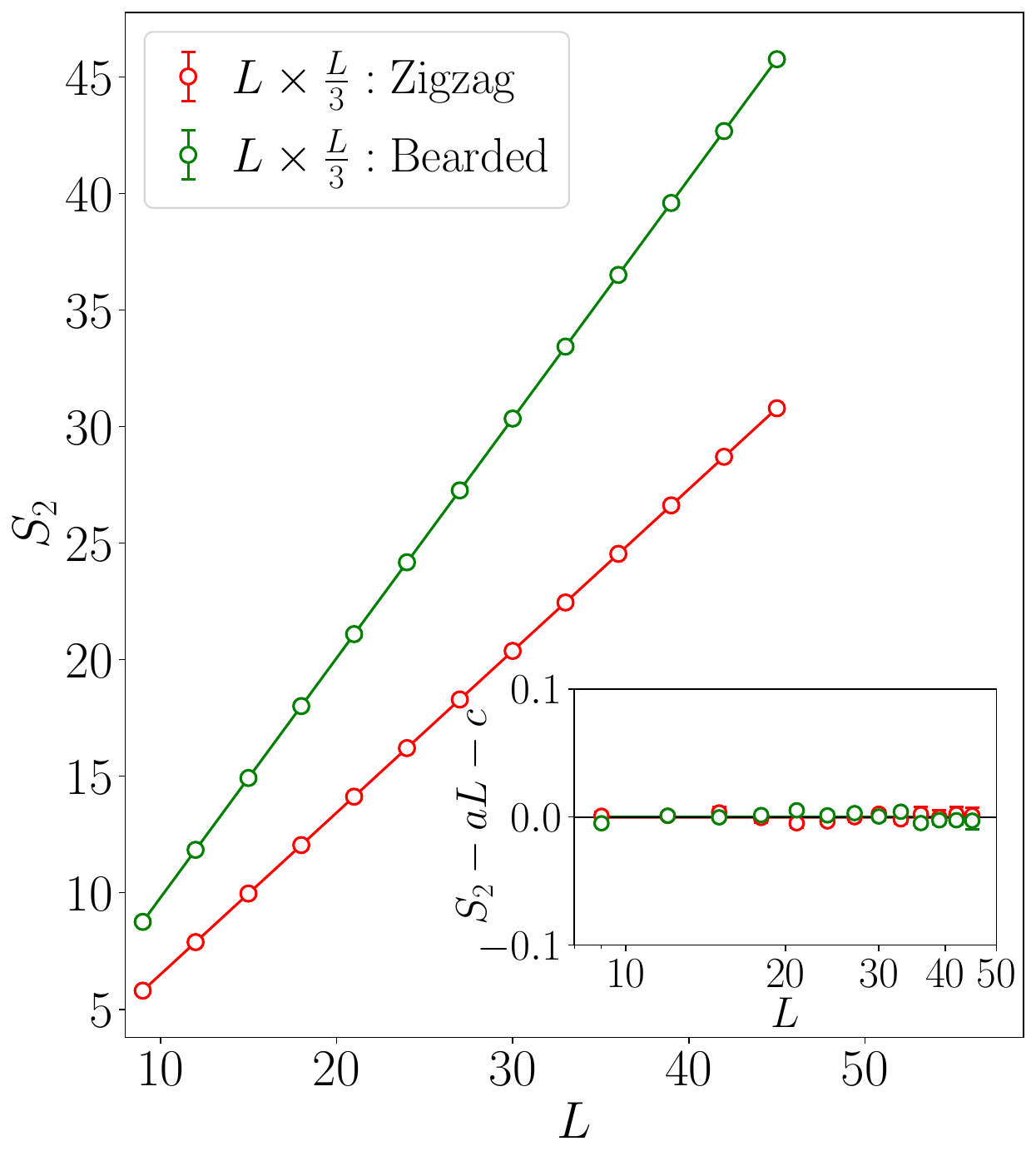}
\put(-2,95){(e)}
\end{overpic}
\end{minipage}
\caption{\label{fig:smooth_L3}
(a--d) Two types of smooth bipartitions of size $L \times L/3$ illustrated on honeycomb lattices with $L=6$ and $L=9$: (a,b) zigzag edges for all $L$, and (c,d) bearded edges for all $L$.
We follow this nomenclature for the edge types in the rest of the figures.
(e) Measured EE for both types of smooth bipartitions (red: zigzag, green: bearded). 
All data have been fitted to the CFT scaling form in Eq.~\ref{eq:2dCFT_form} (solid lines). 
Inset: EE after subtraction of the area-law and constant terms as obtained in our fits plotted versus $\log L$. 
The fits yield $b_{\rm zigzag} = 0.000(9)$ and $b_{\rm bearded} = 0.000(6)$ for the zigzag and bearded edged smooth bipartition, respectively, demonstrating the absence of logarithmic corrections and also showing that this result is insensitive to the edge type.
}
\end{figure*}

\section{Results}
\label{sec:results}

\subsection{Smooth bipartitions}
\label{subsec:smooth}

We begin by studying the scaling of the second R\'enyi entanglement entropy (EE) for smooth bipartitions, for which the EE is expected to obey a pure area law without additional logarithmic corrections from CFT considerations.
We restrict the system size $L$ to multiples of three to ensure commensurability with the Kekul\'e or columnar valence-bond-solid (VBS) pattern on the honeycomb lattice.
Several choices of smooth bipartition are possible, such as subsystems of size $L \times L/2$ or $L \times L/3$ embedded within an $L \times L$ lattice.
Here we focus on the $L \times L/3$ bipartition.
The reason for this choice is that, on the honeycomb lattice, such a construction naturally admits two distinct classes of smooth boundaries:
one in which both subsystem boundaries are of ``zigzag'' type, and another in which both boundaries are of ``bearded'' type.
Representative examples of such zigzag-edge and bearded-edge bipartitions are shown in Fig.~\ref{fig:smooth_L3}(a,b) and Fig.~\ref{fig:smooth_L3}(c,d), respectively, for $6 \times 6$ (even $L$) and $9 \times 9$ (odd $L$) lattices.

The EE data for the zigzag (red) and bearded (green) smooth bipartitions are shown in Fig.~\ref{fig:smooth_L3}(e). We fit the data to the CFT scaling form in Eq.~\ref{eq:2dCFT_form}, including a possible logarithmic term.
In the inset of Fig.~\ref{fig:smooth_L3}(e), we plot the EE after subtracting the fitted area-law and constant terms. The fits yield $b_{\rm zigzag} = 0.000(9)$ and $b_{\rm bearded} = 0.000(6)$ for the zigzag and bearded cases, respectively. 
For both types of smooth bipartitions, EE follows the area law with no logarithmic corrections, as expected from CFT.
This also rules out additional logarithmic terms that would arise from Goldstone modes had the QCP on the honeycomb lattice been a coexistence point.
We have also studied other subsystem choices with mixed zigzag-bearded boundary structure for completeness apart from the zigzag-zigzag and bearded-bearded cases discussed above, and found no logarithmic corrections. On general grounds, this difference from the boundary structure should be inconsequential for large subsystems.


\subsection{Bipartitions with corners}
\label{subsec:corners}


\begin{figure}[t]
\begin{overpic}[width=0.48\linewidth,clip=true,trim=0 0 0 2cm]{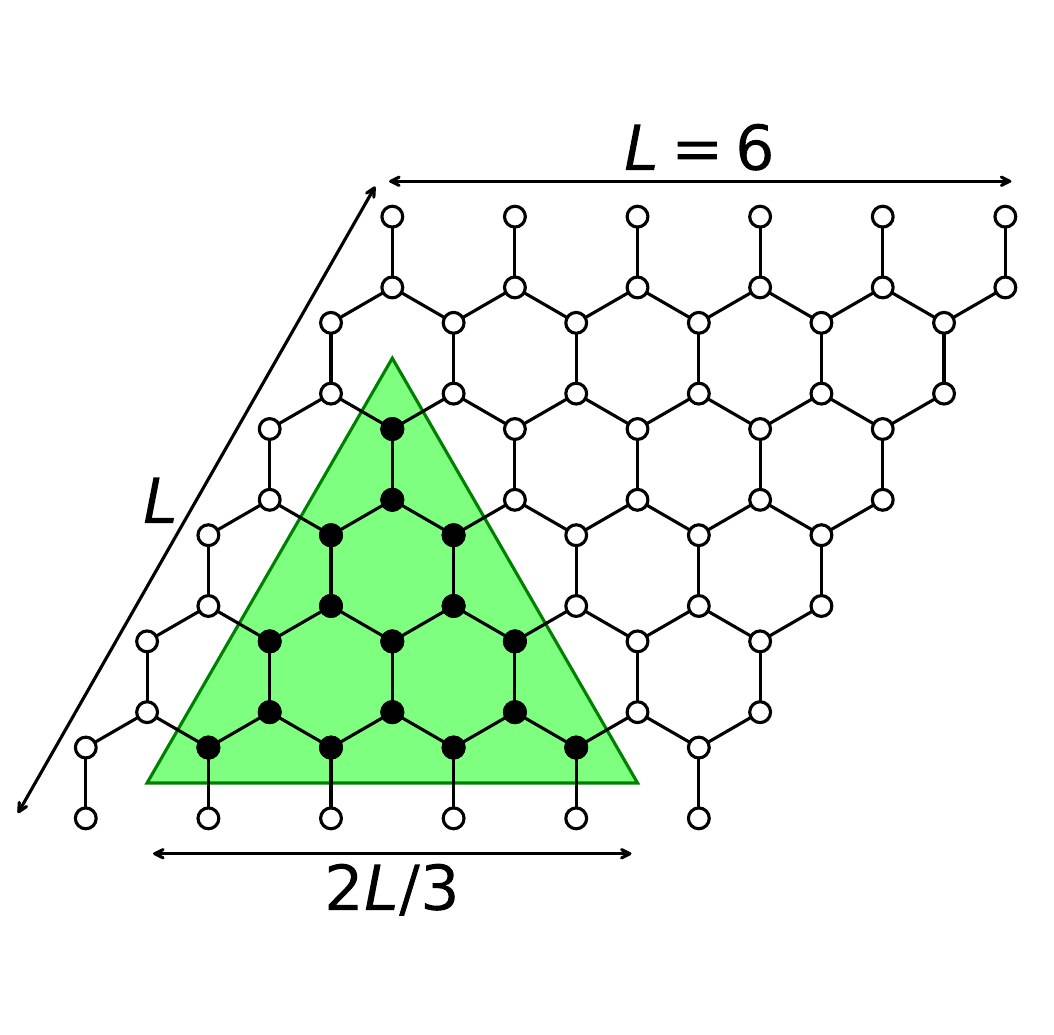}
\put(4,80){(a)}
\end{overpic}
\begin{overpic}[width=0.48\linewidth,clip=true,trim=0 0 0 2cm]{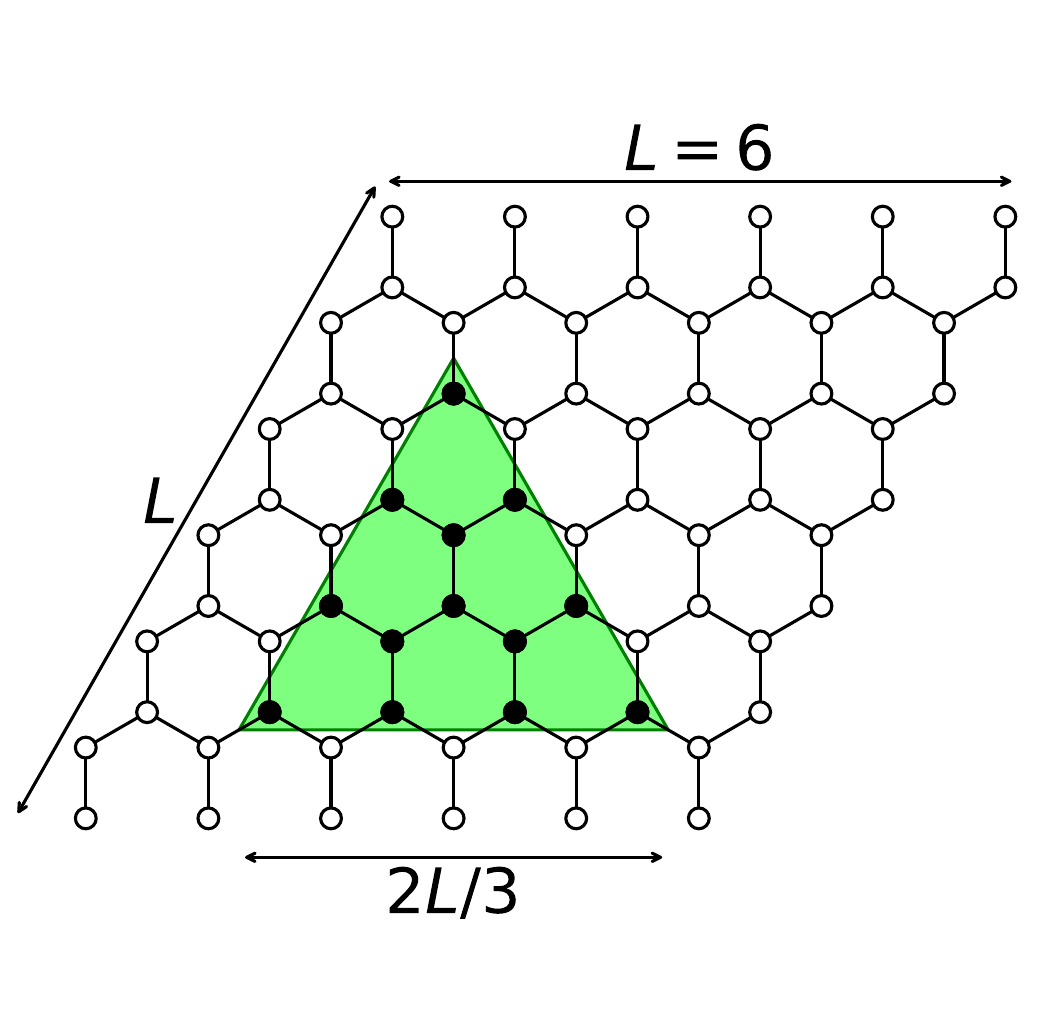}
\put(2,80){(b)}
\end{overpic}
\begin{overpic}[width=0.99\linewidth,clip=true,trim=0 0 0 0.2cm]
{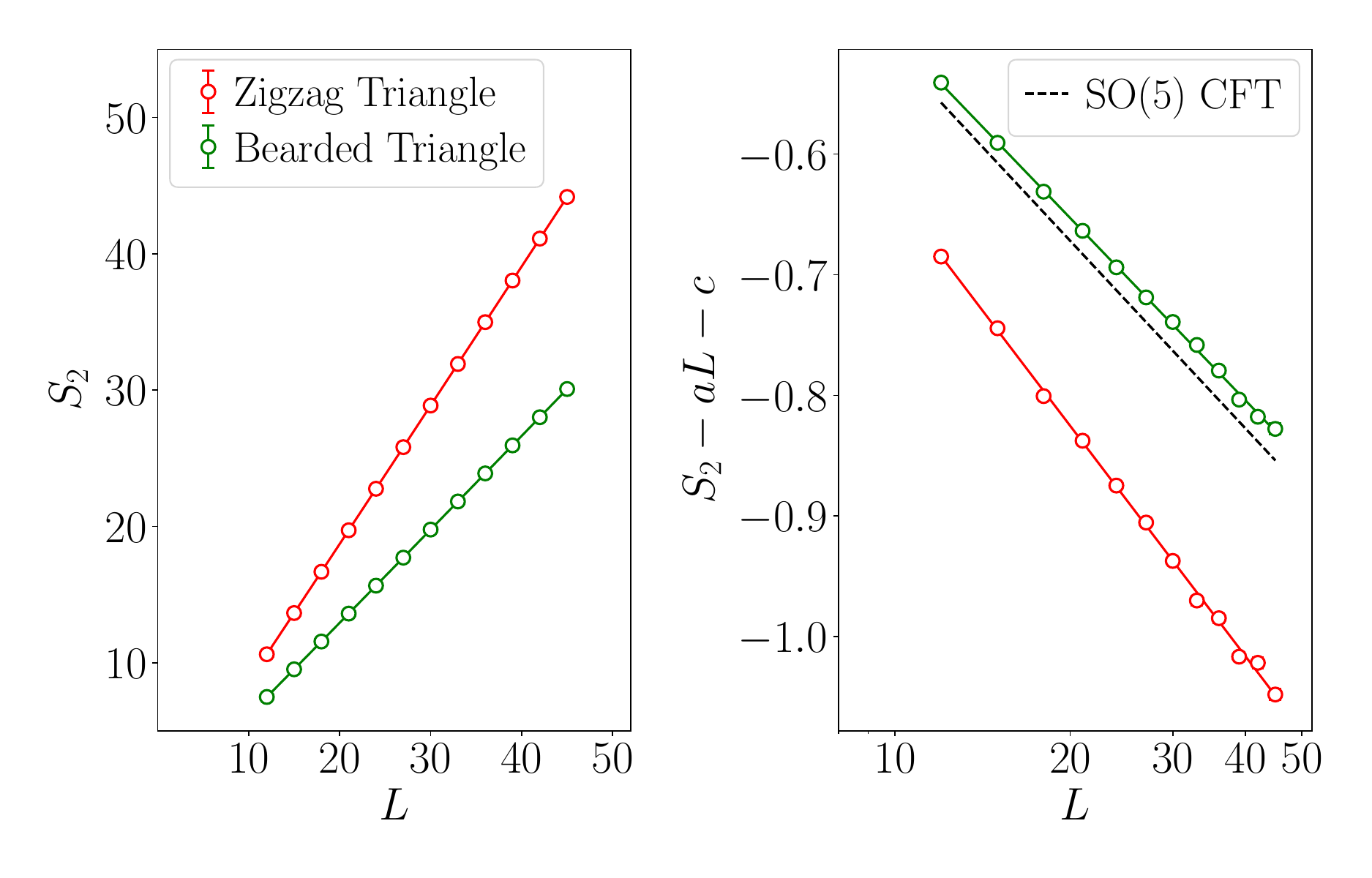}
\put(1,60){(c)}
\put(50,60){(d)}
\end{overpic}
\caption{ \label{fig:triangular}
(a,b) Triangular bipartitions of size $2L/3$ illustrated on a honeycomb lattice with $L=6$: (a) all edges of zigzag type and (b) all edges of bearded type.
(c) Measured EE data for zigzag (red points) and bearded (green points) triangular bipartitions, fitted separately to the CFT scaling form in Eq.~\ref{eq:2dCFT_form} (solid lines). 
(d) The same data after subtracting the fitted area-law and constant terms and plotted versus $\log L$. 
The fits yield $b_{60^\circ} = 0.091(3)$ for the zigzag case and $b_{60^\circ} = 0.072(3)$ for the bearded case. 
The bearded result is in excellent agreement with the 5-component prediction $b_{60^\circ} = 0.074$. 
The dotted line shows the expected slope using the 5-component value as a guide. 
The zigzag case also shows agreement with the 5-component value after using a subtraction scheme; see Appendix \ref{apdx: subtraction}.
}
\end{figure}


\begin{figure*}[t]
\begin{overpic}[width=0.25\linewidth]{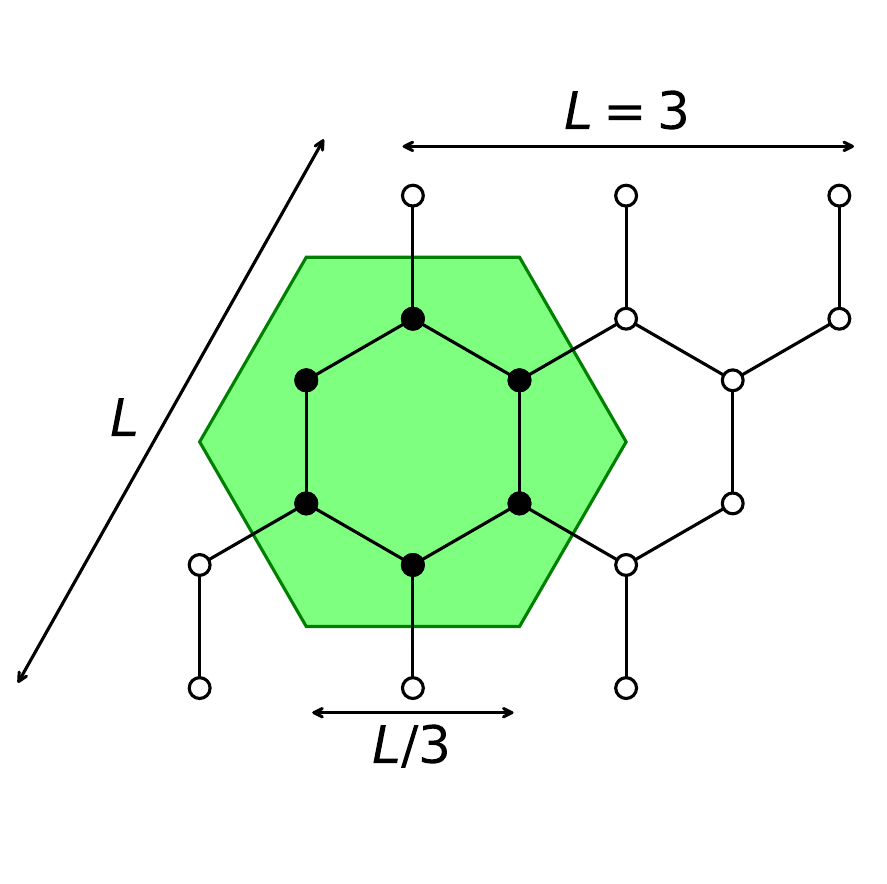}
\put(4,85){(a)}
\end{overpic}
\begin{overpic}[width=0.25\linewidth]{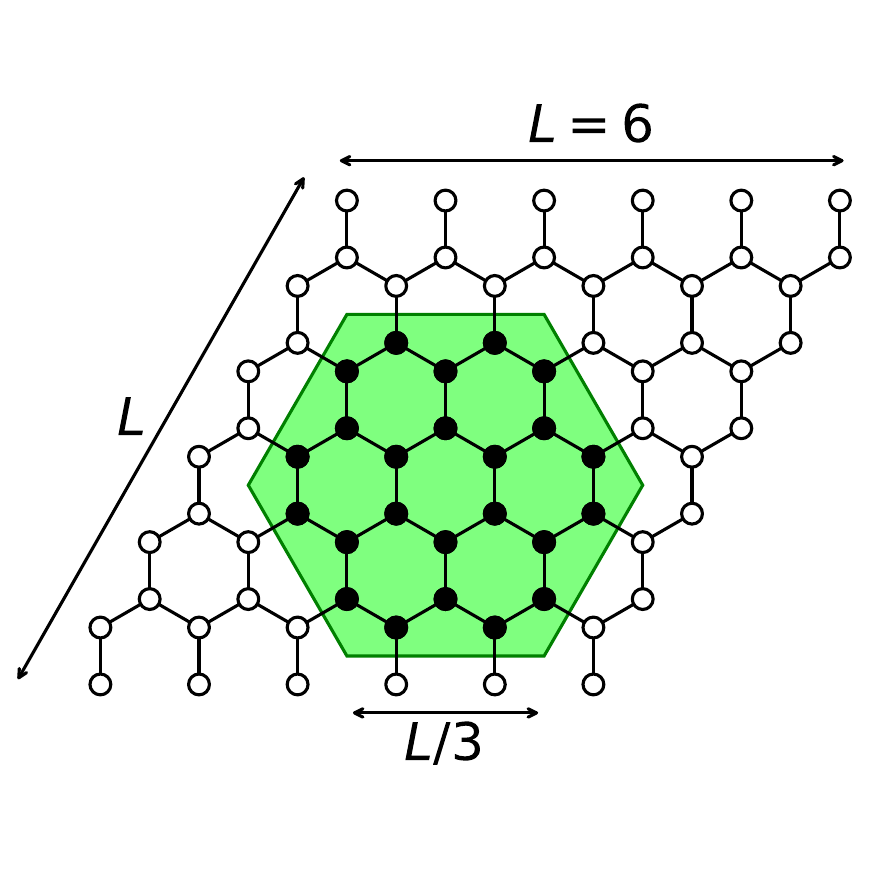}
\put(4,85){(b)}
\end{overpic}
\begin{overpic}[width=0.25\linewidth]{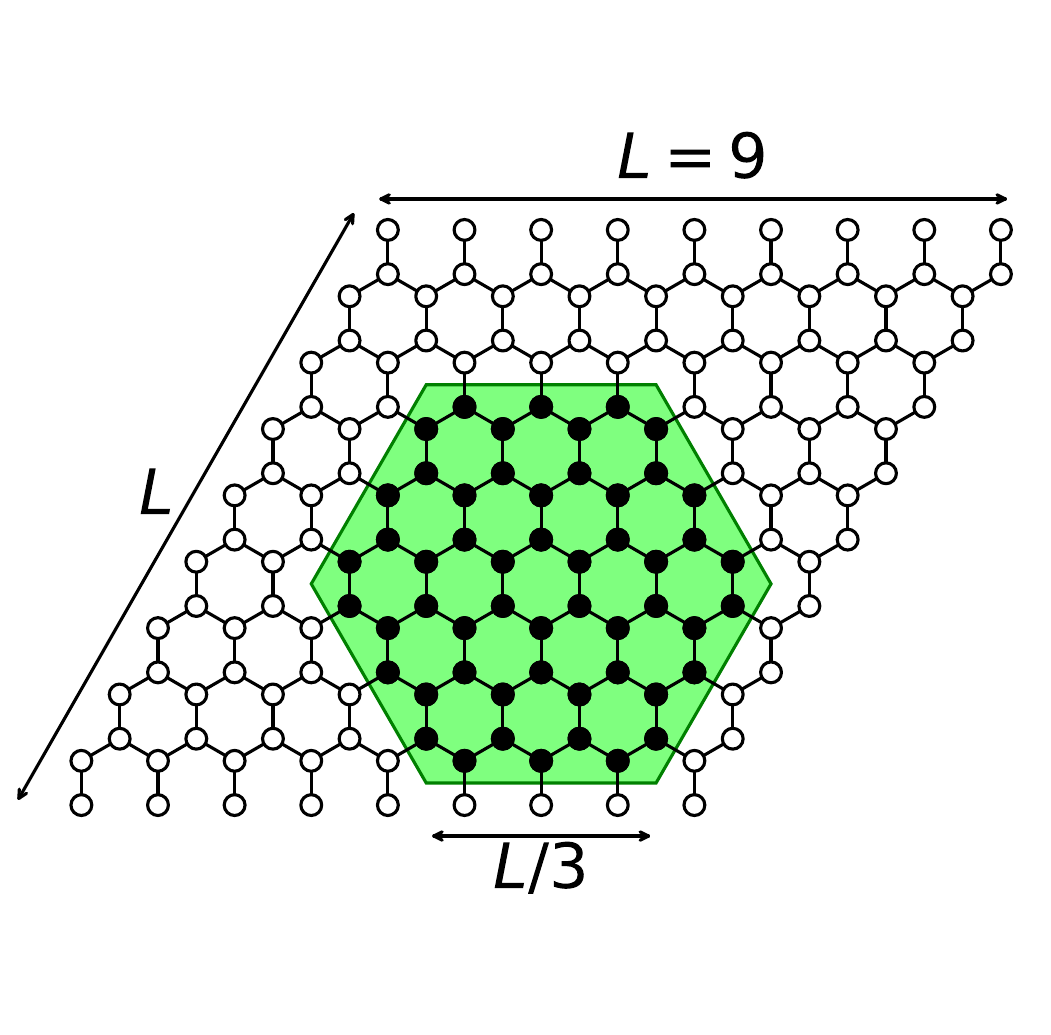}
\put(4,85){(c)}
\end{overpic}
\begin{overpic}[width=0.33\linewidth,clip=true,trim=0 0 0 0cm]{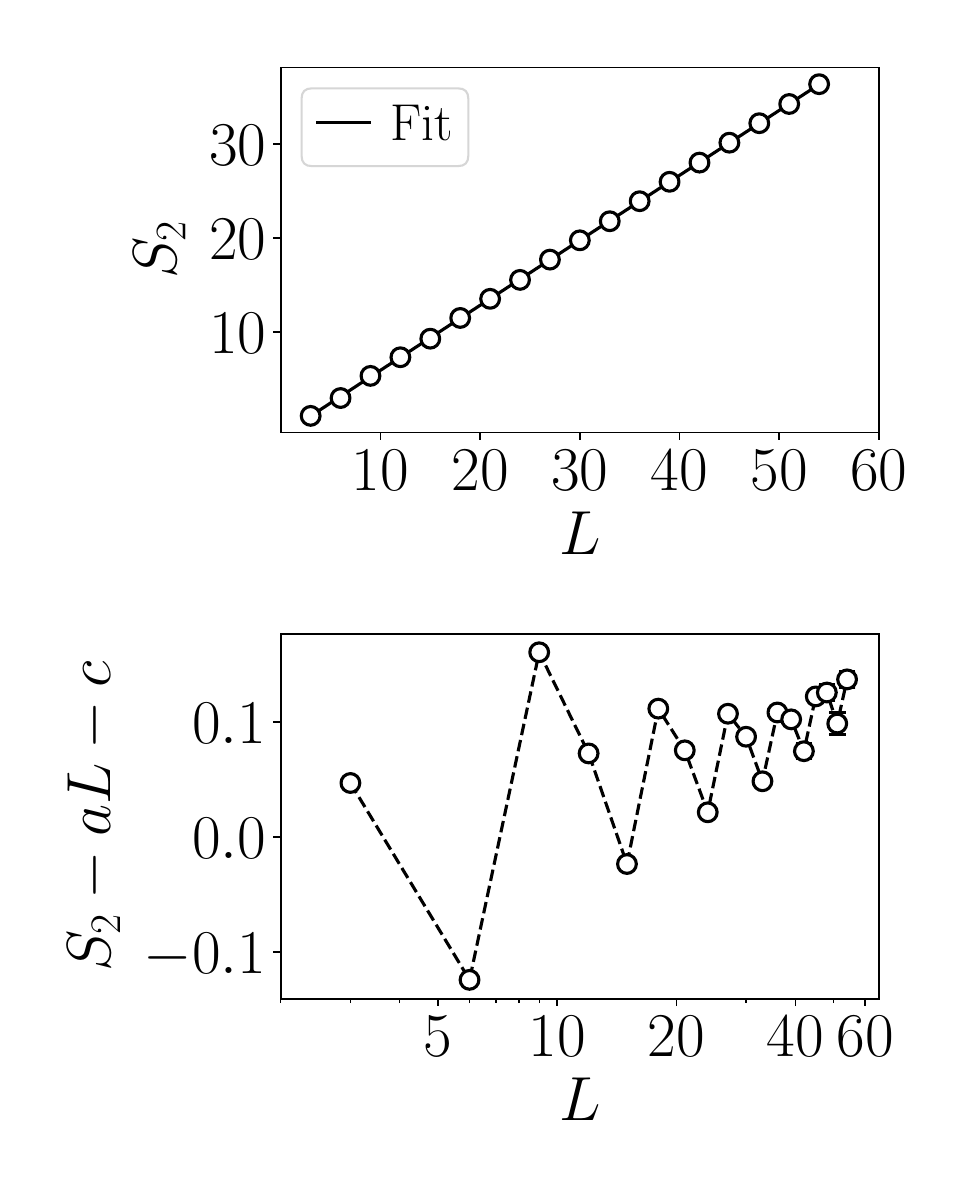}
\put(1,95){(d)}
\put(1,48){(e)}
\end{overpic}
\begin{overpic}[width=0.33\linewidth,clip=true,trim=0 0 0 0cm]{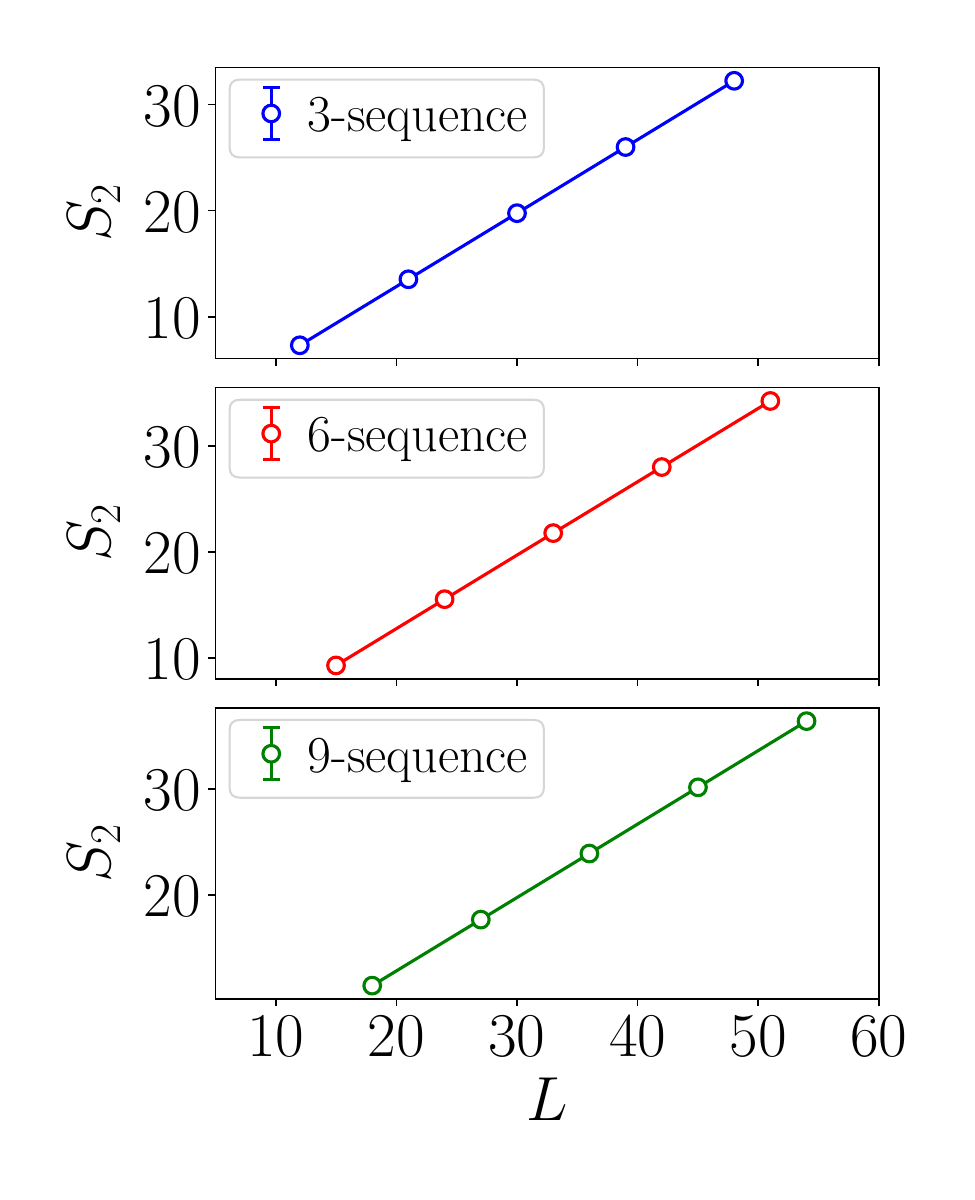}
\put(1,95){(f)}
\end{overpic}
\begin{overpic}[width=0.33\linewidth,clip=true,trim=0 0 0 0cm]{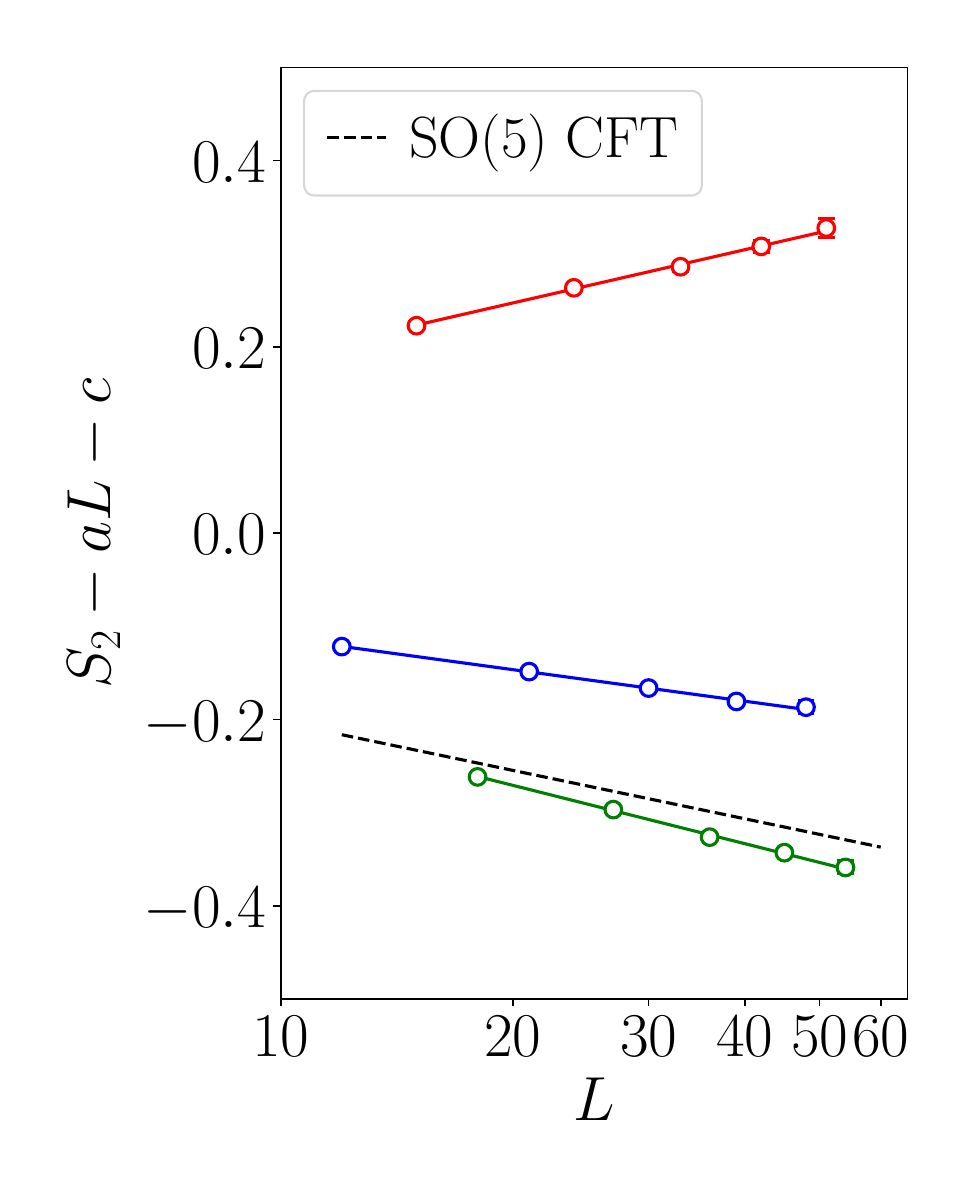}
\put(1,95){(g)}
\end{overpic}
\caption{\label{fig:hexagonal}
(a--c)Hexagonal bipartition of size $L/3$ illustrated on honeycomb lattices with $L=3,6,9$. All edges of the bipartition are of zigzag type and remain unchanged as $L$ varies.
(d) Measured EE data (points) together with fits to the CFT scaling form in Eq.~\ref{eq:2dCFT_form} (solid lines).
(e) The same EE data after subtracting the area-law and constant contributions obtained from the fits, plotted on a log scale in $L$. The subtracted data does not exhibit the expected purely logarithmic behavior; instead, it shows an approximate period-3 oscillation. This demonstrates that the full data set cannot be consistently fitted to the scaling form.
(f) EE data grouped into three system-size sequences, $L=9m+\{3,6,9\}$ with $m\in\{0,1,2,\ldots\}$. Points denote the measured EE data, while solid lines show separate fits of each sequence to the CFT scaling form in Eq.~\ref{eq:2dCFT_form}.
(g) Sequence-separated data after subtracting the fitted area-law and constant terms, plotted on a log scale in $L$. All three sequences now exhibit clear logarithmic behavior, demonstrating successful fits. The extracted coefficients are $b_{120^\circ}=0.008(2)$ (3-sequence), $-0.013(4)$ (6-sequence), and $0.015(4)$ (9-sequence). Only the 9-sequence agrees well with the 5-component value $b=0.0125$. The dotted line denotes this expected 5-component slope.
}
\end{figure*}

Having established that smooth bipartitions show no logarithmic corrections in the previous section, we now turn to bipartitions with sharp corners where additional universal contributions are expected.
As discussed in the introduction, CFT predicts such subsystems to acquire logarithmic corrections in the subsystem EE in presence of corners.
The coefficients of these logarithmic terms depend only on the corner angles and the relevant CFT. 
We will now investigate this feature in the honeycomb $J$-$Q$ model.

We begin with a triangular bipartition of linear size $2L/3$, featuring three $60^\circ$ corners as shown in Fig.~\ref{fig:triangular}.
On the honeycomb lattice, we take triangular bipartitions with two different boundary types for all values of $L$ in this study: one in which all three edges are of zigzag type, and another in which all three edges are of bearded type. 
In Fig.~\ref{fig:triangular}(a,b), we illustrate the bearded and zigzag triangular bipartitions, respectively, on a $6 \times 6$ lattice.
The EE results for the zigzag and bearded triangular bipartitions are shown in Fig.~\ref{fig:triangular}(c), along with fits to the CFT scaling form in Eq.~\ref{eq:2dCFT_form}. 
In Fig.~\ref{fig:triangular}(d), we show the same data after subtracting the fitted area-law and constant terms.
For the zigzag triangular bipartition, we obtain $b_{60^\circ} = 0.091(3)$, while for the bearded triangular bipartition, we obtain $b_{60^\circ} = 0.072(3)$.
The extracted corner coefficient is positive in both cases which is in agreement with the CFT criterion on the positivity of corner coefficients.
We also find that the zigzag corner coefficient is larger than that of the bearded case.
Most notably, the bearded subsystem yields $b_{60^\circ} = 0.072(3)$, remarkably close to the 5-component value~\cite{Whitsitt_etal_2017}, which we reiterate is taken to be five times the Gaussian scalar field theory value $b_{60^\circ} = 0.074$ ~\cite{Helmes_etal_2016} as per the large-$N$ caveat made in Sec.~\ref{sec:intro}. 
We note that such a difference in the corner coefficients between zigzag and bearded triangular cuts has also been reported previously in an entirely different setting -- the honeycomb Hubbard model which hosts a QCP between a Dirac semi-metal and a N\'eel antiferromagnetic Mott insulator, described by the Gross-Neveu-Yukawa fixed point \cite{Jon_etal_Hubbard_2024}.

One possibility to explain such differences could be additional edge-related logarithmic contributions for subsystems with corners. To address this, a specific subtraction method involving two subsystems was devised in the supplement of Ref.~\cite{Jon_sandvik_Properbipartition_2024} to nullify such potential additional contributions. 
We have also performed such an analysis for subsystems with corners which, it turns out, does resolve the above observed difference for $b_{60^\circ}$. 
Since this method involves two subsystems, it is a different analysis when compared to the entanglement scaling behavior of a single bipartitioned subsystem.
We restrict our focus in the main text to single bipartitions and report the subtraction method based results in an appendix (App.~\ref{apdx: subtraction}).


Next, we consider a hexagonal bipartition of linear size $L/3$, featuring six $120^\circ$ corners.
In Fig.~\ref{fig:hexagonal}(a--c), we illustrate this bipartition on $3 \times 3$, $6 \times 6$, and $9 \times 9$ lattices, respectively, showing that all edges are of zigzag type for all system sizes $L$ in this study.
In Fig.~\ref{fig:hexagonal}(d), the raw $S_2$ data is plotted as a function of $L$. $S_2$ clearly scales with $L$ as expected due to the leading area law.
When we try to fit it to the CFT scaling form, we find that it cannot be consistently fitted to it when all system sizes are considered together.
This becomes evident in Fig.~\ref{fig:hexagonal}(e), where we plot the EE after subtracting the area-law contributions and constant terms as obtained from the attempted fit.
The subtracted data do not exhibit the expected pure logarithmic behavior; instead, they show a period-3 oscillation.
This indicates that although the subleading logarithmic contribution is small compared to the leading area-law term, making the raw EE appear smooth, it is not uniform across all system sizes.

Upon closer inspection, we find that the data separates into three distinct sequences depending on $L \bmod 9$: the $3$-sequence ($L = 3, 12, 21, 30, 39, \dots$), the $6$-sequence ($L = 6, 15, 24, 33, 42, \dots$), and the $9$-sequence ($L = 9, 18, 27, 36, 45, \dots$). 
When analyzed independently, each sequence is well described by the linear plus logarithmic scaling form.
In Fig.~\ref{fig:hexagonal}(f), we show the EE for these sequences along with the corresponding fits, and in Fig.~\ref{fig:hexagonal}(g), the same EE data after subtracting the fitted area-law contributions and constant terms. 
In contrast to the combined data set (Fig.~\ref{fig:hexagonal}(e)), the individual sequence data set shows clear linear behavior across the three sequences, albeit with different logarithmic coefficients.
Notably, for the $9$-sequence we obtain $b_{120^\circ} = 0.015(4)$ which is in good agreement with the 5-component value $b_{120^\circ} = 0.0125$
~\cite{Helmes_etal_2016}. 
For the $3$- and $6$-sequences, we obtain $b_{120^\circ}= 0.008(2)$ and $b_{120^\circ} = -0.013(4)$ respectively which do not match with the 5-component value, and the 6-sequence apparently violates the CFT positivity criterion on $b$.


\section{Discussion on proper bipartition and surface criticality arguments}
\label{sec:more_discussion}

In this section, we will discuss our results in light of recent arguments put forth in the literature to understand entanglement scaling of single bipartitioned subsystems near a DQCP.
We recall from Sec.~\ref{sec:intro} that EE scaling at the DQCP in square-lattice $J$-$Q$ models have been shown to depend sensitively on the bipartitioning.
In particular, it was observed that tilted  bipartitions yield EE consistent with $\mathrm{SO}(5)$ CFT expectations~\cite{Jon_sandvik_Properbipartition_2024}.
The tilting refers to the bipartition edges lying at $45^\circ$ angle with respect to the primitive lattice vectors of the square lattice.
Whereas the ``untilted" bipartitions with edges parallel to the primitive lattice vectors show deviations including violations of the positivity of corner coefficients~\cite{Zhao_etal_2022}. 
This led to the notion of \emph{proper bipartitioning}, introduced to identify geometries that correctly capture the universal CFT data.
Subsequent work~\cite{Zhu_surface_criticality} suggested that these differences may be related to the presence of ``surface criticality" at the bipartition edges.
In particular, it was argued that additional gapless edge modes in certain geometries can contribute to the EE, potentially modifying the extracted bulk quantities.
In light of all this, it is natural to examine our results for the honeycomb $J$-$Q$ model from the perspectives of proper bipartitioning and surface criticality.
We find that our observations do not show a clear or systematic consistency with either of these expectations.
While some features may be qualitatively compatible, the overall behavior does not follow the trends suggested by either the proper bipartitioning picture or the surface criticality arguments.
We are not attempting to draw firm conclusions either way in the subsequent discussion, which is outside the scope of this study and left to the future. 

\begin{table*}[t]
\centering
\caption{\label{tab:summary_b} Summary of bipartitions studied in this work, their boundary types, extracted logarithmic coefficients $b$, and the corresponding CFT predictions including the square lattice ones from the literature. The theory values are the leading order result as obtained from a 5-component ($N=5$) Gaussian free scalar field theory ~\cite{Whitsitt_etal_2017,Helmes_etal_2016}. See the large-$N$ caveat in Sec.~\ref{sec:intro} for a justification to use the leading order Gaussian values to compare with the numerically obtained values.}
\renewcommand{\arraystretch}{1.5}
\begin{tabular}{|l|c|c|c|c|}
\hline\hline
Bipartition (on $L \times L$ honeycomb lattice) & Geometry & Edge type & $b_\theta$ (QMC) & $b_\theta$ (CFT) \\
\hline
Smooth : $L \times L/3$ & Strip & zigzag & $0.000(9)$ & $0$ \\
Smooth : $L \times L/3$ & Strip & bearded & $0.000(6)$ & $0$ \\
Triangular : $2L/3$ & $3 \times 60^\circ$ & zigzag & $0.091(3)$ & $0.074$ \\
Triangular : $2L/3$ & $3 \times 60^\circ$ & bearded & $0.072(3)$ & $0.074$ \\
Hexagonal : $L/3$ (3-sequence) & $6 \times 120^\circ$ & zigzag & $0.008(2)$ & $0.0125$ \\
Hexagonal : $L/3$ (6-sequence) & $6 \times 120^\circ$ & zigzag & $-0.013(4)$ & $0.0125$ \\
Hexagonal  : $L/3$ (9-sequence) & $6 \times 120^\circ$ & zigzag & $0.015(4)$ & $0.0125$  \\
\hline\hline
Bipartition (on $L \times L$ square lattice) & Geometry & Bipartition type & $b_\theta$ (QMC) & $b_\theta$ (CFT) \\
\hline
Smooth : $L \times L/2$ & Strip & straight cut & $-0.22(2)$ & $0$ \\
Smooth : $L \times L/2$ & Strip & tilted cut & $0.00(2)$ & $0$  \\
square : $L/2 \times L/2$ &  $4 \times 90^\circ$ & straight cut & $-0.057(2)$ & $0.032$   \\
square : $L/2 \times L/2$ &  $4 \times 90^\circ$ & tilted cut & $0.032(1)$ & $0.032$   \\
\hline\hline
\end{tabular}
\end{table*}

We begin with proper bipartitioning which stipulates that the bipartition boundary treat all the VBS patterns on an equal footing~\cite{Jon_sandvik_Properbipartition_2024}.
The smooth bipartitions and both zigzag and bearded triangular bipartitions from Sec.~\ref{subsec:smooth},~\ref{subsec:corners} satisfy this criterion of proper bipartitioning.
For the smooth bipartitions, we found no logarithmic corrections, while the triangular bipartitions   always yielded positive corner coefficients as required by CFT. 
Agreement with the 5-component value was however only seen for the bearded case and not for the zigzag case even though both cases satisfy the proper bipartitioning criterion. 
We note here that the true value of the corner coefficient at the DQCP is not known, and the 5-component Gaussian scalar field value has been simply used throughout as a point of reference (large-$N$ caveat in Sec.~\ref{sec:intro}), although all of this is highly suggestive of an emergent SO(5) symmetry. 
The more important fact is that the bearded and zigzag triangles do not agree in the corner coefficient when seeing through the proper bipartitioning lens.
Also the hexagonal subsystem is quite interesting from the perspective of proper bipartitioning.
Among the three sequences, \emph{only} the 9-sequence satisfies proper bipartitioning, while the 3- and 6-sequences \emph{do not}.
Consistently, we found that only the 9-sequence yields a corner coefficient in agreement with the CFT prediction.
In this sense, the proper bipartitioning picture provides a coherent explanation of hexagonal subsystem observations.

We next examine whether surface criticality plays a decisive role in our results.
For the smooth bipartitions, we considered geometries with different types of edges in Sec.~\ref{subsec:smooth}.
Although the detailed characterization of surface critical behavior of these edges would require its own separate study (e.g. using the entanglement spectrum based methods of Ref.~\cite{Zhu_surface_criticality}), our results show that any such effects, if present, do not play a dominant role. 
In particular, if additional gapless edge modes were to contribute appreciably, one would expect logarithmic corrections to the area law.
However, for all smooth bipartitions we found no such corrections, with the coefficient of the logarithmic contribution being zero within statistical error showing that they are effectively negligible.
A more striking observation arises from the hexagonal bipartition.
In this case, the EE data separates into three distinct sequences (3-, 6-, and 9-sequences), even though all edges are of the same (zigzag) type for all system sizes $L$.
Since the nature of the edges does not change with $L$, this behavior is difficult to reconcile with a surface criticality picture alone.
The emergence of distinct sequences with different scaling despite identical boundary geometries suggests that surface critical effects are not the primary factor governing the observed EE scaling behavior.

In summary, while the proper bipartitioning criterion seems to account for several of our observations, it does not provide a complete explanation.
Similarly, surface criticality does not seem to offer an explanation for the EE behavior for the mysterious case of sequence separation in case of hexagonal bipartitions.
These results indicate that additional factors beyond proper bipartitioning and surface criticality are perhaps at play in the entanglement scaling at the honeycomb DQCP.
It is furthermore worth asking how the above considerations differ between the honeycomb and the square lattice geometries.
It is an open question if either of the two scenarios above can somehow be reconciled with the set of honeycomb results reported here, or perhaps there exists an altogether different scenario that explains results on both square and honeycomb DQCPs together.


\section{Conclusion}
\label{sec:conclusion}

In this work, we investigated in detail the scaling of the second-order R\'enyi entanglement entropy at the N\'eel-VBS DQCP on the honeycomb lattice using a sign-free $J$-$Q$ model (Eq.~\ref{eq:JQ3_model} in Sec.~\ref{sec:model}) and large-scale QMC simulations (Sec.~\ref{sec:QMC_method}). 
By considering a variety of subsystem geometries, it is found that the EE scaling of the DQCP is consistent with unitary CFT based predictions over a range of analysis set-ups as summarized in Table~\ref{tab:summary_b}. 
In other words, the suspected weakly first-order nature of the transition in this model does not seem to play a significant role in the EE scaling.
In summary, for smooth bipartitions we observed no logarithmic corrections to the area law independent of edge geometry in agreement with CFT expectations.
For subsystems with corners, such as triangular and hexagonal regions that exhibit critical scaling behavior given by Eq.~\ref{eq:2dCFT_form}, we were able to extract  corner coefficients that are consistent with a 5-component scalar field order parameter, highly suggestive of an emergent $\mathrm{SO}(5)$ CFT.
Finally, we also examined our results from the recent perspectives of proper bipartitioning criterion and surface criticality in Sec.~\ref{sec:more_discussion} and found gaps in both cases when trying to explain our full set of results.

Recent results from a fuzzy sphere analysis~\cite{Zhou_etal_PRX_2024} based on a set-up that is different from $J$-$Q$ models had come to an apparently opposite conclusion that ``$\ldots$ the Néel-VBS transition on the honeycomb lattice cannot be described by the SO(5) DQCP''. 
This was based on identifying the RG-relevance of tripled monopoles  through the fuzzy sphere method supporting a pseudo-critical or approximately conformal scenario.
As background, see Sec.~\ref{sec:model} for discussion related to this point.
However, Ref.~\cite{Zhou_etal_PRX_2024} also concluded a stable DQCP on the square lattice based on the RG-irrelevance of fourfold monopoles, while the square lattice $J$-$Q$ model Néel-VBS transition is now believed to be weakly first-order.
Given the remarkably good critical behavior seen in our honeycomb results for rather large subsystems, it is natural to ask if our results and the fuzzy sphere results are at odds needing a resolution, or whether the approximately conformal scenario is a sufficient explanation to reconcile the two results.
Recall that Ref.~\cite{Pujari_Damle_Alet_2013,Pujari_Damle_Alet_2015} had come to the conclusion that tripled monopoles were (dangerously) irrelevant in the $J$-$Q$ honeycomb model based on order-parameter scaling results.
An analysis based on Ref.~\cite{demidio2021diagnosing} on the honeycomb lattice would thus be quite illuminating in this regard.
From a field theory perspective, one may speculate on the question of which operator(s) dominantly control the order parameter and entanglement based scaling, and if this could somehow be an ``in-principle'' source of differences in their scaling behaviors.

\vspace{0.15cm}
Overall, the EE scaling results on the honeycomb lattice along with the existing square lattice results in the literature form a collection of evidence that shows the universality of the N\'eel-VBS DQCP with an emergent SO(5) CFT that is, to a very good extent, accessible in lattice $J$-$Q$ models.
However, all these results also show that the boundary structure of the bipartitions is playing an important role in the EE scaling behavior.
Table ~\ref{tab:summary_b} summarizes them including the square lattice ones from the literature. 
Thus EE scaling studies from QMC simulations, apart from verifying field-theoretic predictions for the DQCP scenario, are also serving up new puzzles for field theory including the intriguing period-3 oscillation reported here.


\vspace{0.5cm}
\section*{Acknowledgements}
We acknowledge computational resource of the PARAM Rudra and Chandra HPC clusters.
SP acknowledges support from ANRF-DST (formerly SERB), Govt. of India via Grant No. MTR/2022/000386 and partially by Grant No. CRG/2021/003024.
S.P. also acknowledges the Indo-Japan LOTUS science exchange award for partial support during the final stages of this project. J. D. acknowledges support from the National Science Foundation (NSF) under Award No. OSI-2326801.

\nocite{Vijigiri2026}
\bibliography{refs}

@article{Shao_Guo_Sandvik_Science_2016,
  title = {Quantum criticality with two length scales},
  volume = {352},
  ISSN = {1095-9203},
  url = {http://dx.doi.org/10.1126/science.aad5007},
  DOI = {10.1126/science.aad5007},
  number = {6282},
  journal = {Science},
  publisher = {American Association for the Advancement of Science (AAAS)},
  author = {Shao,  Hui and Guo,  Wenan and Sandvik,  Anders W.},
  year = {2016},
  month = apr,
  pages = {213–216}
}

@article{Zhou_etal_PRX_2024,
  title = {SO(5) Deconfined Phase Transition under the Fuzzy-Sphere Microscope: Approximate Conformal Symmetry,  Pseudo-Criticality,  and Operator Spectrum},
  volume = {14},
  ISSN = {2160-3308},
  url = {http://dx.doi.org/10.1103/PhysRevX.14.021044},
  DOI = {10.1103/physrevx.14.021044},
  number = {2},
  journal = {Physical Review X},
  publisher = {American Physical Society (APS)},
  author = {Zhou,  Zheng and Hu,  Liangdong and Zhu,  W. and He,  Yin-Chen},
  year = {2024},
  month = june 
}

@article{Senthil_etal_science_2004,
author = {T. Senthil  and Ashvin Vishwanath  and Leon Balents  and Subir Sachdev  and Matthew P. A. Fisher },
title = {Deconfined Quantum Critical Points},
journal = {Science},
volume = {303},
number = {5663},
pages = {1490-1494},
year = {2004},
doi = {10.1126/science.1091806},
URL = {https://www.science.org/doi/abs/10.1126/science.1091806},
eprint = {https://www.science.org/doi/pdf/10.1126/science.1091806}}

@article{Senthil_etal_PRB_2004,
  title = {Quantum criticality beyond the Landau-Ginzburg-Wilson paradigm},
  author = {Senthil, T. and Balents, Leon and Sachdev, Subir and Vishwanath, Ashvin and Fisher, Matthew P. A.},
  journal = {Phys. Rev. B},
  volume = {70},
  issue = {14},
  pages = {144407},
  numpages = {33},
  year = {2004},
  month = {Oct},
  publisher = {American Physical Society},
  doi = {10.1103/PhysRevB.70.144407},
  url = {https://link.aps.org/doi/10.1103/PhysRevB.70.144407}
}

@article{Levin_Senthil_2004,
  title = {Deconfined quantum criticality and N\'eel order via dimer disorder},
  author = {Levin, Michael and Senthil, T.},
  journal = {Phys. Rev. B},
  volume = {70},
  issue = {22},
  pages = {220403},
  numpages = {4},
  year = {2004},
  month = {Dec},
  publisher = {American Physical Society},
  doi = {10.1103/PhysRevB.70.220403},
  url = {https://link.aps.org/doi/10.1103/PhysRevB.70.220403}
}

@article{Tanaka_Hu_2005,
  title = {Many-Body Spin Berry Phases Emerging from the $\ensuremath{\pi}$-Flux State: Competition between Antiferromagnetism and the Valence-Bond-Solid State},
  author = {Tanaka, Akihiro and Hu, Xiao},
  journal = {Phys. Rev. Lett.},
  volume = {95},
  issue = {3},
  pages = {036402},
  numpages = {4},
  year = {2005},
  month = {Jul},
  publisher = {American Physical Society},
  doi = {10.1103/PhysRevLett.95.036402},
  url = {https://link.aps.org/doi/10.1103/PhysRevLett.95.036402}
}

@article{Senthil_Fisher_2006,
  title = {Competing orders,  nonlinear sigma models,  and topological terms in quantum magnets},
  volume = {74},
  ISSN = {1550-235X},
  url = {http://dx.doi.org/10.1103/PhysRevB.74.064405},
  DOI = {10.1103/physrevb.74.064405},
  number = {6},
  journal = {Physical Review B},
  publisher = {American Physical Society (APS)},
  author = {Senthil,  T. and Fisher,  Matthew P. A.},
  year = {2006},
  month = aug 
}

@book{Sachdev_QPTbook_2011, place={Cambridge}, edition={2}, title={Quantum Phase Transitions}, publisher={Cambridge University Press}, author={Sachdev, Subir}, year={2011}}

@article{Sandvik_2007,
  title = {Evidence for Deconfined Quantum Criticality in a Two-Dimensional Heisenberg Model with Four-Spin Interactions},
  author = {Sandvik, Anders W.},
  journal = {Phys. Rev. Lett.},
  volume = {98},
  issue = {22},
  pages = {227202},
  numpages = {4},
  year = {2007},
  month = {Jun},
  publisher = {American Physical Society},
  doi = {10.1103/PhysRevLett.98.227202},
  url = {https://link.aps.org/doi/10.1103/PhysRevLett.98.227202}
}

@article{Pujari_Damle_Alet_2013,
  title = {N\'eel-State to Valence-Bond-Solid Transition on the Honeycomb Lattice: Evidence for Deconfined Criticality},
  author = {Pujari, Sumiran and Damle, Kedar and Alet, Fabien},
  journal = {Phys. Rev. Lett.},
  volume = {111},
  issue = {8},
  pages = {087203},
  numpages = {5},
  year = {2013},
  month = {Aug},
  publisher = {American Physical Society},
  doi = {10.1103/PhysRevLett.111.087203},
  url = {https://link.aps.org/doi/10.1103/PhysRevLett.111.087203}
}

@article{Pujari_Damle_Alet_2015,
  title = {Transitions to valence-bond solid order in a honeycomb lattice antiferromagnet},
  author = {Pujari, Sumiran and Alet, Fabien and Damle, Kedar},
  journal = {Phys. Rev. B},
  volume = {91},
  issue = {10},
  pages = {104411},
  numpages = {14},
  year = {2015},
  month = {Mar},
  publisher = {American Physical Society},
  doi = {10.1103/PhysRevB.91.104411},
  url = {https://link.aps.org/doi/10.1103/PhysRevB.91.104411}
}

@article{Fradkin_etal_2006,
  title = {Entanglement Entropy of 2D Conformal Quantum Critical Points: Hearing the Shape of a Quantum Drum},
  author = {Fradkin, Eduardo and Moore, Joel E.},
  journal = {Phys. Rev. Lett.},
  volume = {97},
  issue = {5},
  pages = {050404},
  numpages = {4},
  year = {2006},
  month = {Aug},
  publisher = {American Physical Society},
  doi = {10.1103/PhysRevLett.97.050404},
  url = {https://link.aps.org/doi/10.1103/PhysRevLett.97.050404}
}

@article{Casini_Huerta_2007,
title = {Universal terms for the entanglement entropy in 2+1 dimensions},
journal = {Nuclear Physics B},
volume = {764},
number = {3},
pages = {183-201},
year = {2007},
issn = {0550-3213},
doi = {https://doi.org/10.1016/j.nuclphysb.2006.12.012},
url = {https://www.sciencedirect.com/science/article/pii/S0550321306010091},
author = {H. Casini and M. Huerta}
}

@article{Whitsitt_etal_2017,
  title = {Entanglement entropy of large-$N$ Wilson-Fisher conformal field theory},
  author = {Whitsitt, Seth and Witczak-Krempa, William and Sachdev, Subir},
  journal = {Phys. Rev. B},
  volume = {95},
  issue = {4},
  pages = {045148},
  numpages = {9},
  year = {2017},
  month = {Jan},
  publisher = {American Physical Society},
  doi = {10.1103/PhysRevB.95.045148},
  url = {https://link.aps.org/doi/10.1103/PhysRevB.95.045148}
}

@article{Helmes_etal_2016,
  title = {Universal corner entanglement of Dirac fermions and gapless bosons from the continuum to the lattice},
  author = {Helmes, Johannes and Hayward Sierens, Lauren E. and Chandran, Anushya and Witczak-Krempa, William and Melko, Roger G.},
  journal = {Phys. Rev. B},
  volume = {94},
  issue = {12},
  pages = {125142},
  numpages = {14},
  year = {2016},
  month = {Sep},
  publisher = {American Physical Society},
  doi = {10.1103/PhysRevB.94.125142},
  url = {https://link.aps.org/doi/10.1103/PhysRevB.94.125142}
}

@article{sandvik_book,
    author = {Sandvik, Anders W.},
    title = {Computational Studies of Quantum Spin Systems},
    journal = {AIP Conference Proceedings},
    volume = {1297},
    number = {1},
    pages = {135-338},
    year = {2010},
    month = {11},
    issn = {0094-243X},
    doi = {10.1063/1.3518900},
    url = {https://doi.org/10.1063/1.3518900},
}

@article{Calabrese_Cardy_2004,
doi = {10.1088/1742-5468/2004/06/P06002},
url = {https://doi.org/10.1088/1742-5468/2004/06/P06002},
year = {2004},
month = {jun},
publisher = {},
volume = {2004},
number = {06},
pages = {P06002},
author = {Pasquale Calabrese and John Cardy},
title = {Entanglement entropy and quantum field theory},
journal = {Journal of Statistical Mechanics: Theory and Experiment}
}

@article{Melko_Kallin_Hasting_2010,
  title = {Finite-size scaling of mutual information in Monte Carlo simulations: Application to the spin-$\frac{1}{2}$ $XXZ$ model},
  author = {Melko, Roger G. and Kallin, Ann B. and Hastings, Matthew B.},
  journal = {Phys. Rev. B},
  volume = {82},
  issue = {10},
  pages = {100409},
  numpages = {4},
  year = {2010},
  month = {Sep},
  publisher = {American Physical Society},
  doi = {10.1103/PhysRevB.82.100409},
  url = {https://link.aps.org/doi/10.1103/PhysRevB.82.100409}
}

@article{Hastings_Gonzalez_Kallin_Melko_2010,
  title = {Measuring Renyi Entanglement Entropy in Quantum Monte Carlo Simulations},
  author = {Hastings, Matthew B. and Gonz\'alez, Iv\'an and Kallin, Ann B. and Melko, Roger G.},
  journal = {Phys. Rev. Lett.},
  volume = {104},
  issue = {15},
  pages = {157201},
  numpages = {4},
  year = {2010},
  month = {Apr},
  publisher = {American Physical Society},
  doi = {10.1103/PhysRevLett.104.157201},
  url = {https://link.aps.org/doi/10.1103/PhysRevLett.104.157201}
}

@article{Humeniuk_Roscilde_2012,
  title = {Quantum Monte Carlo calculation of entanglement R\'enyi entropies for generic quantum systems},
  author = {Humeniuk, Stephan and Roscilde, Tommaso},
  journal = {Phys. Rev. B},
  volume = {86},
  issue = {23},
  pages = {235116},
  numpages = {8},
  year = {2012},
  month = {Dec},
  publisher = {American Physical Society},
  doi = {10.1103/PhysRevB.86.235116},
  url = {https://link.aps.org/doi/10.1103/PhysRevB.86.235116}
}

@article{Inglis_Melko_2013,
  title = {Wang-Landau method for calculating R\'enyi entropies in finite-temperature quantum Monte Carlo simulations},
  author = {Inglis, Stephen and Melko, Roger G.},
  journal = {Phys. Rev. E},
  volume = {87},
  issue = {1},
  pages = {013306},
  numpages = {8},
  year = {2013},
  month = {Jan},
  publisher = {American Physical Society},
  doi = {10.1103/PhysRevE.87.013306},
  url = {https://link.aps.org/doi/10.1103/PhysRevE.87.013306}
}

@article{Luitz_Plat_Laflorencie_Alet_2014,
  title = {Improving entanglement and thermodynamic R\'enyi entropy measurements in quantum Monte Carlo},
  author = {Luitz, David J. and Plat, Xavier and Laflorencie, Nicolas and Alet, Fabien},
  journal = {Phys. Rev. B},
  volume = {90},
  issue = {12},
  pages = {125105},
  numpages = {13},
  year = {2014},
  month = {Sep},
  publisher = {American Physical Society},
  doi = {10.1103/PhysRevB.90.125105},
  url = {https://link.aps.org/doi/10.1103/PhysRevB.90.125105}
}

@article{Kulchytskyy_etal_2015,
  title = {Detecting Goldstone modes with entanglement entropy},
  author = {Kulchytskyy, Bohdan and Herdman, C. M. and Inglis, Stephen and Melko, Roger G.},
  journal = {Phys. Rev. B},
  volume = {92},
  issue = {11},
  pages = {115146},
  numpages = {11},
  year = {2015},
  month = {Sep},
  publisher = {American Physical Society},
  doi = {10.1103/PhysRevB.92.115146},
  url = {https://link.aps.org/doi/10.1103/PhysRevB.92.115146}
}

@article{Alba_2017,
  title = {Out-of-equilibrium protocol for R\'enyi entropies via the Jarzynski equality},
  author = {Alba, Vincenzo},
  journal = {Phys. Rev. E},
  volume = {95},
  issue = {6},
  pages = {062132},
  numpages = {7},
  year = {2017},
  month = {Jun},
  publisher = {American Physical Society},
  doi = {10.1103/PhysRevE.95.062132},
  url = {https://link.aps.org/doi/10.1103/PhysRevE.95.062132}
}

@article{Jon_PRL_2020,
  title = {Entanglement Entropy from Nonequilibrium Work},
  author = {D'Emidio, Jonathan},
  journal = {Phys. Rev. Lett.},
  volume = {124},
  issue = {11},
  pages = {110602},
  numpages = {5},
  year = {2020},
  month = {Mar},
  publisher = {American Physical Society},
  doi = {10.1103/PhysRevLett.124.110602},
  url = {https://link.aps.org/doi/10.1103/PhysRevLett.124.110602}
}

@article{Zhao_etal_2022,
  title = {Scaling of Entanglement Entropy at Deconfined Quantum Criticality},
  author = {Zhao, Jiarui and Wang, Yan-Cheng and Yan, Zheng and Cheng, Meng and Meng, Zi Yang},
  journal = {Phys. Rev. Lett.},
  volume = {128},
  issue = {1},
  pages = {010601},
  numpages = {6},
  year = {2022},
  month = {Jan},
  publisher = {American Physical Society},
  doi = {10.1103/PhysRevLett.128.010601},
  url = {https://link.aps.org/doi/10.1103/PhysRevLett.128.010601}
}

@article{Jon_sandvik_Properbipartition_2024,
  title = {Entanglement Entropy and Deconfined Criticality: Emergent SO(5) Symmetry and Proper Lattice Bipartition},
  author = {D'Emidio, Jonathan and Sandvik, Anders W.},
  journal = {Phys. Rev. Lett.},
  volume = {133},
  issue = {16},
  pages = {166702},
  numpages = {7},
  year = {2024},
  month = {Oct},
  publisher = {American Physical Society},
  doi = {10.1103/PhysRevLett.133.166702},
  url = {https://link.aps.org/doi/10.1103/PhysRevLett.133.166702}
}

@article{Jon_etal_Hubbard_2024,
  title = {Universal Features of Entanglement Entropy in the Honeycomb Hubbard Model},
  author = {D'Emidio, Jonathan and Or\'us, Rom\'an and Laflorencie, Nicolas and de Juan, Fernando},
  journal = {Phys. Rev. Lett.},
  volume = {132},
  issue = {7},
  pages = {076502},
  numpages = {6},
  year = {2024},
  month = {Feb},
  publisher = {American Physical Society},
  doi = {10.1103/PhysRevLett.132.076502},
  url = {https://link.aps.org/doi/10.1103/PhysRevLett.132.076502}
}

@article{Zhu_surface_criticality,
  title = {Bipartite Entanglement and Surface Criticality: The Extra Contribution of the Nonordinary Edge in Entanglement},
  author = {Zhu, Yanzhang and Liu, Zenan and Wang, Zhe and Wang, Yan-Cheng and Yan, Zheng},
  journal = {Phys. Rev. Lett.},
  volume = {136},
  issue = {4},
  pages = {046501},
  numpages = {6},
  year = {2026},
  month = {Jan},
  publisher = {American Physical Society},
  doi = {10.1103/tkx5-kzhh},
  url = {https://link.aps.org/doi/10.1103/tkx5-kzhh}
}

@misc{boundary_effects,
note={For example, see Ref.~\cite{Zhu_surface_criticality} for a quantitative discussion of such boundary effects which we also discuss in Sec.~\ref{sec:more_discussion}.}
}

@article{Lou2009:AFtoVBSsun,
  title = {{Antiferromagnetic to valence-bond-solid transitions in two-dimensional $\text{SU}(N)$ Heisenberg models with multispin interactions}},
  author = {Lou, Jie and Sandvik, Anders W. and Kawashima, Naoki},
  journal = {Phys. Rev. B},
  volume = {80},
  issue = {18},
  pages = {180414},
  numpages = {4},
  year = {2009},
  month = {Nov},
  publisher = {American Physical Society},
  doi = {10.1103/PhysRevB.80.180414},
  url = {https://link.aps.org/doi/10.1103/PhysRevB.80.180414}
}

@article{Sandvik2010:ContinuousJQlog,
  title = {{Continuous Quantum Phase Transition between an Antiferromagnet and a Valence-Bond Solid in Two Dimensions: Evidence for Logarithmic Corrections to Scaling}},
  author = {Sandvik, Anders W.},
  journal = {Phys. Rev. Lett.},
  volume = {104},
  issue = {17},
  pages = {177201},
  numpages = {4},
  year = {2010},
  month = {Apr},
  publisher = {American Physical Society},
  doi = {10.1103/PhysRevLett.104.177201},
  url = {https://link.aps.org/doi/10.1103/PhysRevLett.104.177201}
}

@article{Sandvik2020:Consistent,
	Author = {Anders W. Sandvik and Bowen Zhao},
	Doi = {10.1088/0256-307X/37/5/057502},
	Journal = {Chinese Physics Letters},
	Month = {may},
	Number = {5},
	Pages = {057502},
	Publisher = {Chinese Physical Society and IOP Publishing Ltd},
	Title = {{Consistent Scaling Exponents at the Deconfined Quantum-Critical Point}},
	Url = {https://dx.doi.org/10.1088/0256-307X/37/5/057502},
	Volume = {37},
	Year = {2020}}

@article{Zhao2020:Helical,
  title = {{Multicritical Deconfined Quantum Criticality and Lifshitz Point of a Helical Valence-Bond Phase}},
  author = {Zhao, Bowen and Takahashi, Jun and Sandvik, Anders W.},
  journal = {Phys. Rev. Lett.},
  volume = {125},
  issue = {25},
  pages = {257204},
  numpages = {7},
  year = {2020},
  month = {Dec},
  publisher = {American Physical Society},
  doi = {10.1103/PhysRevLett.125.257204},
  url = {https://link.aps.org/doi/10.1103/PhysRevLett.125.257204}
}

@article{Harada2013:DQCPsmallN,
  title = {{Possibility of deconfined criticality in SU($N$) Heisenberg models at small $N$}},
  author = {Harada, Kenji and Suzuki, Takafumi and Okubo, Tsuyoshi and Matsuo, Haruhiko and Lou, Jie and Watanabe, Hiroshi and Todo, Synge and Kawashima, Naoki},
  journal = {Phys. Rev. B},
  volume = {88},
  issue = {22},
  pages = {220408},
  numpages = {4},
  year = {2013},
  month = {Dec},
  publisher = {American Physical Society},
  doi = {10.1103/PhysRevB.88.220408},
  url = {https://link.aps.org/doi/10.1103/PhysRevB.88.220408}
}

@article{Block2013:Fate,
  title = {{Fate of $\mathbb{C}{\mathbb{P}}^{N\ensuremath{-}1}$ Fixed Points with $q$ Monopoles}},
  author = {Block, Matthew S. and Melko, Roger G. and Kaul, Ribhu K.},
  journal = {Phys. Rev. Lett.},
  volume = {111},
  issue = {13},
  pages = {137202},
  numpages = {5},
  year = {2013},
  month = {Sep},
  publisher = {American Physical Society},
  doi = {10.1103/PhysRevLett.111.137202},
  url = {https://link.aps.org/doi/10.1103/PhysRevLett.111.137202}
}

@article{Kuklov2008:DCPfirstorder,
  title = {{Deconfined Criticality: Generic First-Order Transition in the SU(2) Symmetry Case}},
  author = {Kuklov, A. B. and Matsumoto, M. and Prokof'ev, N. V. and Svistunov, B. V. and Troyer, M.},
  journal = {Phys. Rev. Lett.},
  volume = {101},
  issue = {5},
  pages = {050405},
  numpages = {4},
  year = {2008},
  month = {Aug},
  publisher = {American Physical Society},
  doi = {10.1103/PhysRevLett.101.050405},
  url = {https://link.aps.org/doi/10.1103/PhysRevLett.101.050405}
}

@article{Jiang2008:FirstOrder,
	Author = {F-J Jiang and M Nyfeler and S Chandrasekharan and U-J Wiese
},
	Doi = {10.1088/1742-5468/2008/02/P02009},
	Journal = {Journal of Statistical Mechanics: Theory and Experiment},
	Month = {feb},
	Number = {02},
	Pages = {P02009},
	Title = {{From an antiferromagnet to a valence bond solid: evidence for a first-order phase transition}},
	Url = {https://dx.doi.org/10.1088/1742-5468/2008/02/P02009},
	Volume = {2008},
	Year = {2008},
	Url = {https://dx.doi.org/10.1088/1742-5468/2008/02/P02009}
}

@article{Chen2013:DCPflow,
  title = {{Deconfined Criticality Flow in the Heisenberg Model with Ring-Exchange Interactions}},
  author = {Chen, Kun and Huang, Yuan and Deng, Youjin and Kuklov, A. B. and Prokof'ev, N. V. and Svistunov, B. V.},
  journal = {Phys. Rev. Lett.},
  volume = {110},
  issue = {18},
  pages = {185701},
  numpages = {5},
  year = {2013},
  month = {May},
  publisher = {American Physical Society},
  doi = {10.1103/PhysRevLett.110.185701},
  url = {https://link.aps.org/doi/10.1103/PhysRevLett.110.185701}
}

@Article{demidio2021diagnosing,
	title={{Diagnosing weakly first-order phase transitions by coupling to order parameters}},
	author={Jonathan D'Emidio and Alexander A. Eberharter and Andreas M. Läuchli},
	journal={SciPost Phys.},
	volume={15},
	pages={061},
	year={2023},
	publisher={SciPost},
	doi={10.21468/SciPostPhys.15.2.061},
	url={https://scipost.org/10.21468/SciPostPhys.15.2.061},
}

@misc{Takahashi2024:multicritical,
      title={{SO(5) multicriticality in two-dimensional quantum magnets}}, 
      author={Jun Takahashi and Hui Shao and Bowen Zhao and Wenan Guo and Anders W. Sandvik},
      year={2024},
      eprint={2405.06607},
      archivePrefix={arXiv},
      primaryClass={cond-mat.str-el},
      url={https://arxiv.org/abs/2405.06607}, 
}

@article{Wang2017:DQCPsymm,
  title = {Deconfined Quantum Critical Points: Symmetries and Dualities},
  author = {Wang, Chong and Nahum, Adam and Metlitski, Max A. and Xu, Cenke and Senthil, T.},
  journal = {Phys. Rev. X},
  volume = {7},
  issue = {3},
  pages = {031051},
  numpages = {45},
  year = {2017},
  month = {Sep},
  publisher = {American Physical Society},
  doi = {10.1103/PhysRevX.7.031051},
  url = {https://link.aps.org/doi/10.1103/PhysRevX.7.031051}
}

@article{Ma2020:DQCPpseudo,
  title = {Theory of deconfined pseudocriticality},
  author = {Ma, Ruochen and Wang, Chong},
  journal = {Phys. Rev. B},
  volume = {102},
  issue = {2},
  pages = {020407(R)},
  numpages = {6},
  year = {2020},
  month = {Jul},
  publisher = {American Physical Society},
  doi = {10.1103/PhysRevB.102.020407},
  url = {https://link.aps.org/doi/10.1103/PhysRevB.102.020407}
}

@article{Nahum2020:Note,
  title = {Note on Wess-Zumino-Witten models and quasiuniversality in $2+1$ dimensions},
  author = {Nahum, Adam},
  journal = {Phys. Rev. B},
  volume = {102},
  issue = {20},
  pages = {201116(R)},
  numpages = {4},
  year = {2020},
  month = {Nov},
  publisher = {American Physical Society},
  doi = {10.1103/PhysRevB.102.201116},
  url = {https://link.aps.org/doi/10.1103/PhysRevB.102.201116}
}

@article{Gorbenko2018:Walking,
	Author = {Gorbenko, Victor and Rychkov, Slava and Zan, Bernardo},
	Da = {2018/10/16},
	Doi = {10.1007/JHEP10(2018)108},
	Id = {Gorbenko2018},
	Isbn = {1029-8479},
	Journal = {Journal of High Energy Physics},
	Number = {10},
	Pages = {108},
	Title = {{Walking, weak first-order transitions, and complex CFTs}},
	Ty = {JOUR},
	Url = {https://doi.org/10.1007/JHEP10(2018)108},
	Volume = {2018},
	Year = {2018}}

@Article{Gorbenko2018:Walking2,
	title={{Walking, Weak first-order transitions, and Complex CFTs II.  Two-dimensional Potts model at $Q>4$}},
	author={Victor Gorbenko and Slava Rychkov and Bernardo Zan},
	journal={SciPost Phys.},
	volume={5},
	pages={050},
	year={2018},
	publisher={SciPost},
	doi={10.21468/SciPostPhys.5.5.050},
	url={https://scipost.org/10.21468/SciPostPhys.5.5.050},
}

@article{Kumar2026:PseudoSUN,
  title = {Pseudocriticality in Antiferromagnetic Spin Chains},
  author = {Kumar, Sankalp and Pujari, Sumiran and D'Emidio, Jonathan},
  journal = {Phys. Rev. Lett.},
  volume = {136},
  issue = {7},
  pages = {076701},
  numpages = {7},
  year = {2026},
  month = {Feb},
  publisher = {American Physical Society},
  doi = {10.1103/vlzz-gzy5},
  url = {https://link.aps.org/doi/10.1103/vlzz-gzy5}
}

@article{Song2024:subleadingEE,
	title = {Extracting subleading corrections in entanglement entropy at quantum phase transitions},
	pages = {010},
	author = {Song, Menghan and Zhao, Jiarui and Meng, Zi Yang and Xu, Cenke and Cheng, Meng},
	journal = {SciPost Phys.},
	volume = {17},
	year = {2024},
	publisher = {SciPost},
	doi = {10.21468/SciPostPhys.17.1.010},
	url = {https://scipost.org/10.21468/SciPostPhys.17.1.010}
}

@article{Deng2024:Diagnosing,
  title = {Diagnosing Quantum Phase Transition Order and Deconfined Criticality via Entanglement Entropy},
  author = {Deng, Zehui and Liu, Lu and Guo, Wenan and Lin, Hai-Qing},
  journal = {Phys. Rev. Lett.},
  volume = {133},
  issue = {10},
  pages = {100402},
  numpages = {6},
  year = {2024},
  month = {Sep},
  publisher = {American Physical Society},
  doi = {10.1103/PhysRevLett.133.100402},
  url = {https://link.aps.org/doi/10.1103/PhysRevLett.133.100402}
}

@article{Mao2026:Detecting,
  title = {Detecting the Emergent Continuous Symmetry of Criticality via a Subsystem's Entanglement Spectrum},
  author = {Mao, Bin-Bin and Wang, Zhe and Chen, Bin-Bin and Yan, Zheng},
  journal = {Phys. Rev. Lett.},
  volume = {136},
  issue = {4},
  pages = {046401},
  numpages = {8},
  year = {2026},
  month = {Jan},
  publisher = {American Physical Society},
  doi = {10.1103/7j21-l3pg},
  url = {https://link.aps.org/doi/10.1103/7j21-l3pg}
}

@article{Murthy1990:Hedgehog,
title = {Action of hedgehog instantons in the disordered phase of the (2+1)-dimensional CPN-1 model},
journal = {Nuclear Physics B},
volume = {344},
number = {3},
pages = {557-595},
year = {1990},
issn = {0550-3213},
doi = {https://doi.org/10.1016/0550-3213(90)90670-9},
url = {https://www.sciencedirect.com/science/article/pii/0550321390906709},
author = {Ganpathy Murthy and Subir Sachdev},
}

@article{Vijigiri2026,
  title = {Deconfined pseudocriticality in a model spin-1 quantum antiferromagnet},
  volume = {38},
  ISSN = {1361-648X},
  url = {http://dx.doi.org/10.1088/1361-648X/ae334d},
  DOI = {10.1088/1361-648x/ae334d},
  number = {2},
  journal = {Journal of Physics: Condensed Matter},
  publisher = {IOP Publishing},
  author = {Vijigiri,  Vikas and Pujari,  Sumiran and Desai,  Nisheeta},
  year = {2026},
  month = Jan,
  pages = {025601}
}

@misc{footnote_binder_dips,
    note={However, numerical studies on both honeycomb and square lattices have not found any first-order related ``dips'' in the staggered magnetization Binder ratio. See Ref.~\cite{Vijigiri2026} for more on this point, in particular Secs.~3.4 and 5 therein.}
}



\appendix


\section{Subtraction method}
\label{apdx: subtraction}

\begin{figure}[t]
\begin{overpic}[width=0.48\linewidth]{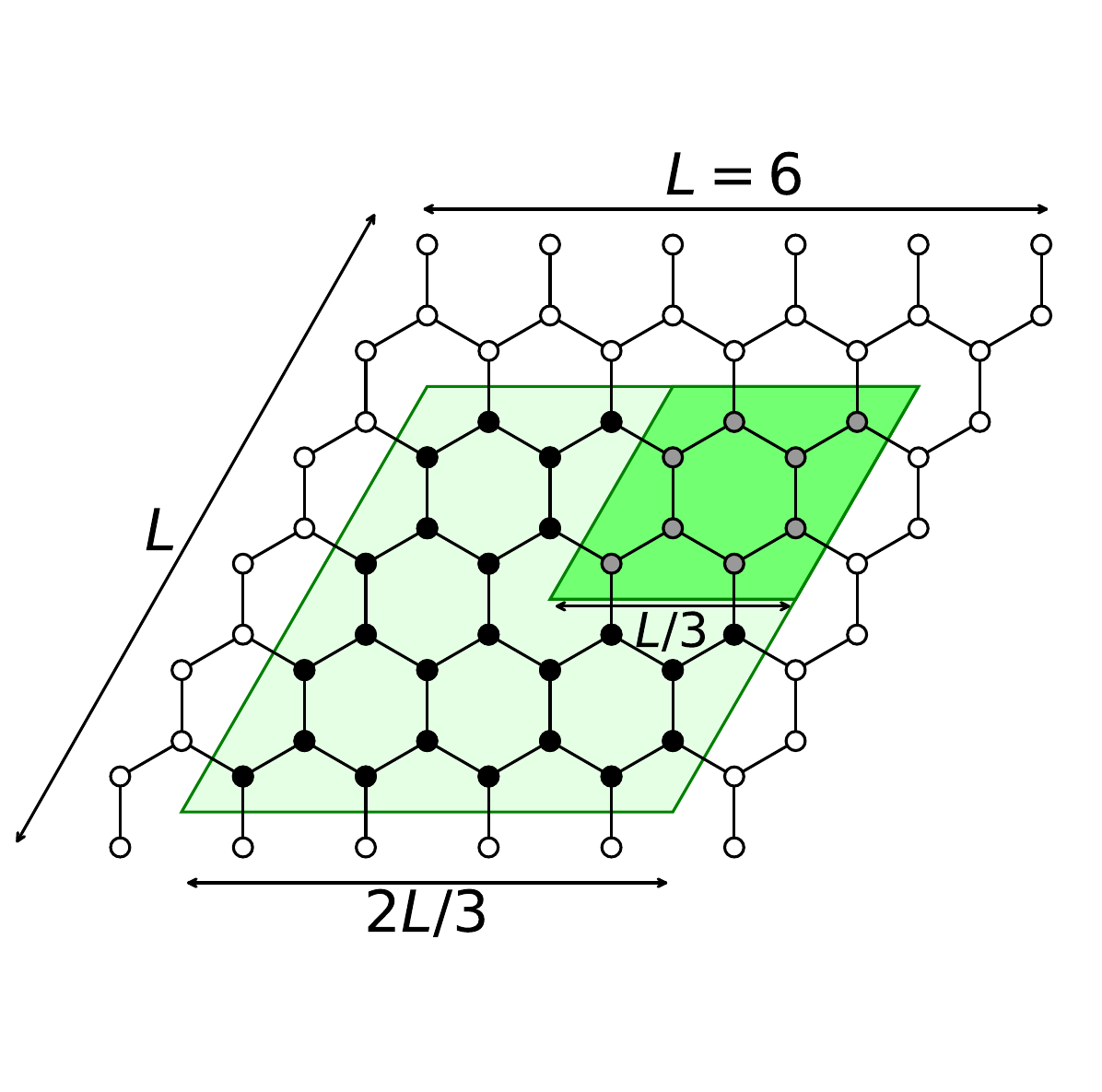}
\put(4,85){(a)}
\end{overpic}
\begin{overpic}[width=0.48\linewidth]{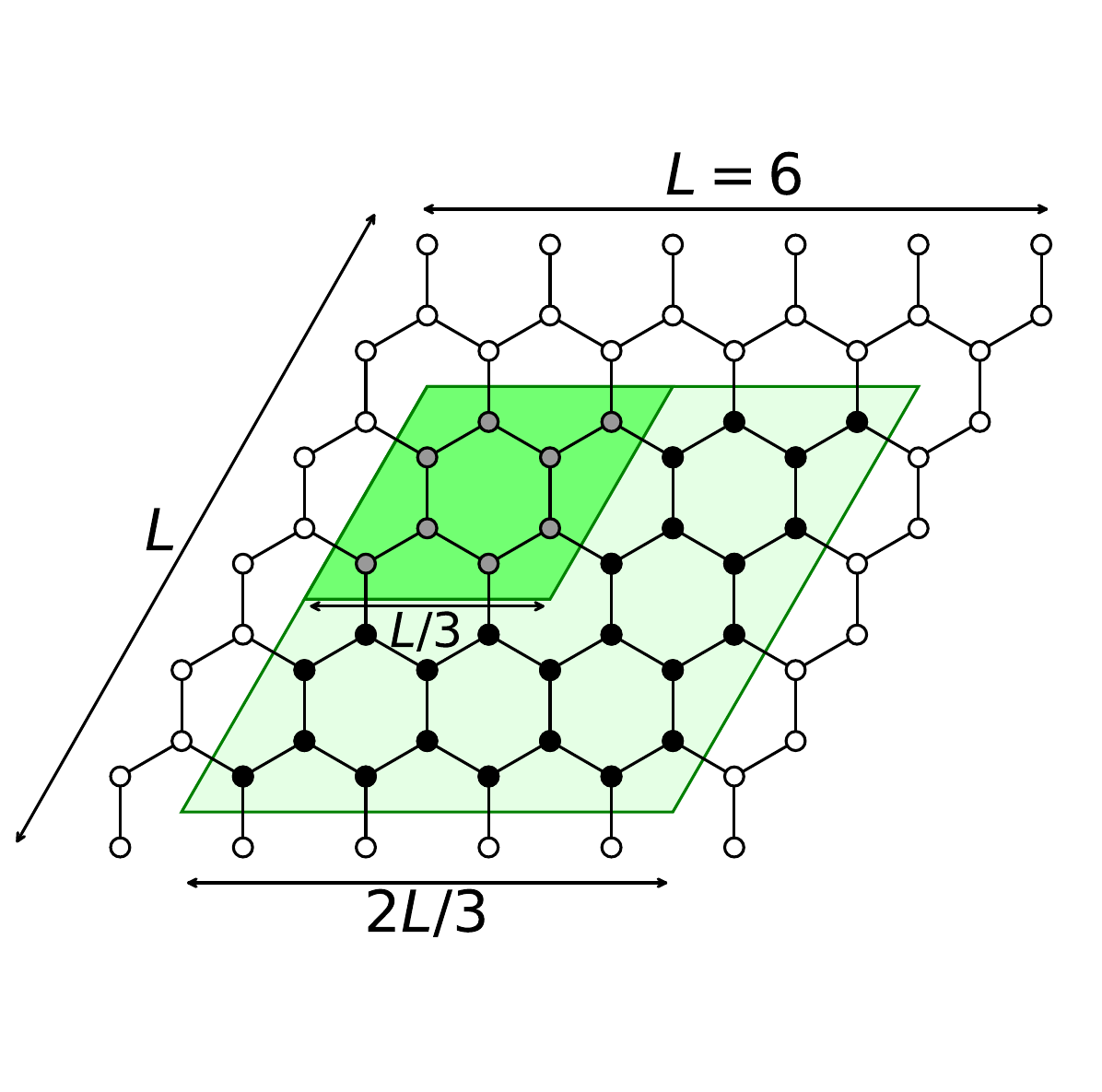}
\put(4,85){(b)}
\end{overpic}
\begin{overpic}[width=0.49\linewidth]{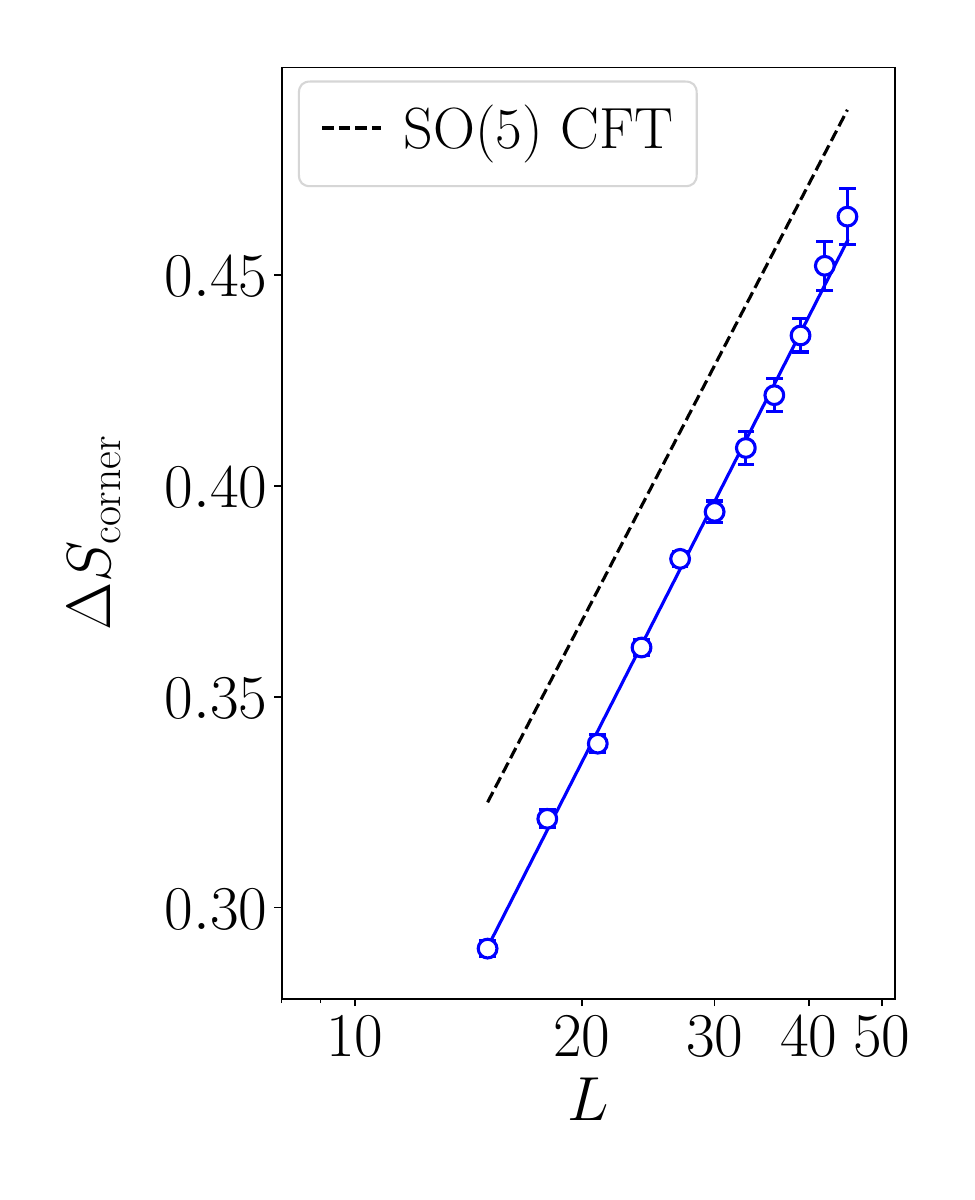}
\put(4,90){(c)}
\end{overpic}
\begin{overpic}[width=0.49\linewidth]{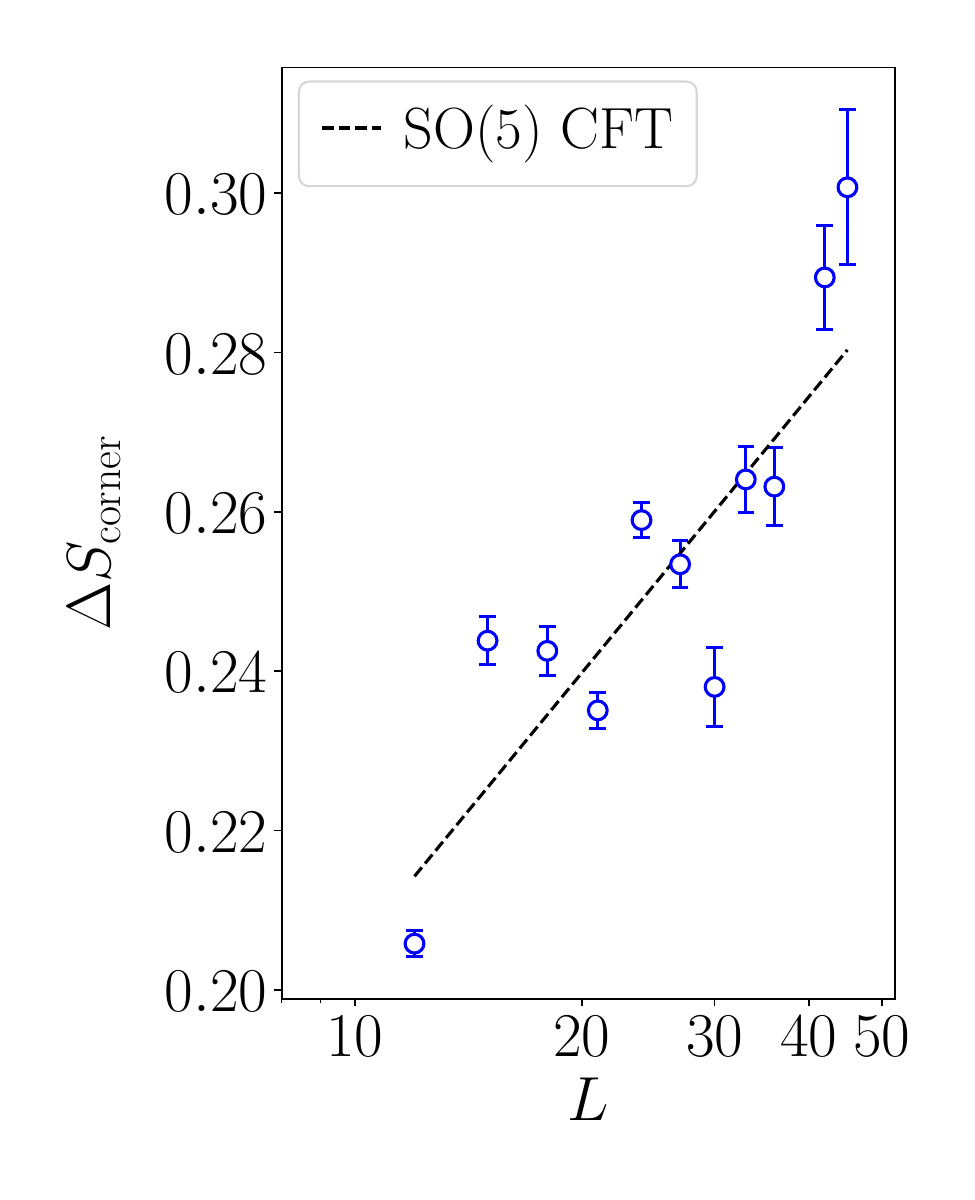}
\put(4,90){(d)}
\end{overpic}
\caption{\label{fig:subtraction}
(a) Subsystems used to isolate the $60^\circ$ corner contribution to the EE  
based on a subtraction scheme shown through the case of $L=6$ system size. 
By subtracting the EE of these two subsystems with identical perimeters, the edge related contributions would cancel out leaving only the contribution from the two $60^\circ$ corners (Eq.~\ref{eq:deltaS_60}).
(b) Subsystems used to isolate the $120^\circ$ corner contribution using an analogous construction.
(c) The subtracted EE $(\Delta S_{\text{corner}})$ for the construction in panel (a) plotted as a function of $\log L$, along with a linear fit used to extract the corner coefficient. 
The fit gives $b_{60^\circ} = 0.076(1)$ which is consistent with the 5-component prediction ($b_{60^\circ} = 0.074$). 
The dotted line indicates the expected slope from the 5-component scenario.
(d) The $\Delta S_{\text{corner}}$ for the $120^\circ$ case as a function of $\log L$. 
In contrast to (c), we find that the data does not exhibit a clear linear trend, and therefore we can not reliably extract the corner coefficient $b_{120^\circ}$ through this method.
The dotted line indicates the expected slope from the 5-component scenario.
}
\end{figure}


From the EE scaling of the triangular and hexagonal bipartitions, we found that agreement with a putative SO(5) CFT seems to sensitively depend on the boundary structure or geometry.
From a continuum perspective, these details may be considered not too relevant.
However, since the SO(5) order parameter is assumed to be obtained by some coarse-graining procedure of the space-time configuration of the micrscopic degrees of freedom, the boundary degrees of freedom can possibly contribute to the bulk EE~\cite{boundary_effects}.
From a phenomenological perspective, we may practically say that perhaps edges introduce additional logarithmic contributions to the EE beyond the universal corner term.
For example in case of the triangular bipartitions, the corner coefficient $b_{60^\circ}$ extracted from the bearded case was found to be in good agreement with the CFT prediction, whereas the zigzag case yielded a larger value which may be phenomenologically attributed to such an additional contribution.

To minimize such potential edge-related contributions, we perform here a specific subtraction scheme developed in the supplement of  Ref. ~\cite{Jon_sandvik_Properbipartition_2024} to investigate further and cross-check with direct fitting of the EE as in the main text (Sec.~\ref{subsec:smooth},~\ref{subsec:corners}).
The idea simply is to consider two subsystems with identical perimeters and the same type of edges, but different numbers or types of corners.
In this way, any additional edge contributions are expected to be identical for both subsystems and cancel upon subtraction. By subtracting their entanglement entropies, the leading area-law term would also cancel along with any potential logarithmic corrections from the edges, leaving only the logarithmic contributions from the corners.
We will call such subtracted EE data by $\Delta S_{\text{corner}}$.

In Fig.~\ref{fig:subtraction}(a), we show one such pair of subsystems used to extract the $60^\circ$ corner coefficient on a $6 \times 6$ lattice. 
The light green shaded subsystem contains four $60^\circ$ and two $120^\circ$ corners, while the combined (light + bright green) subsystem has two $60^\circ$ and two $120^\circ$ corners. 
All edges in both subsystems are of zigzag type. Since both subsystems have the same perimeter and the same number of $120^\circ$ corners, the area-law contribution and the logarithmic terms from the $120^\circ$ corners would cancel. 
We therefore expect the dominant term in the scaling
\begin{equation}
\Delta S_{\text{corner}} \simeq 
2 b_{60^\circ} \ln L ,
\label{eq:deltaS_60}    
\end{equation}
arising from the difference of two $60^\circ$ corners.
Figure~\ref{fig:subtraction}(c) shows $\Delta S_{\text{corner}}$ plotted as a function of $\log L$, along with a fit to the above form up to a possible non-universal and thus unimportant $O(L^0)$ constant.
The extracted value $b_{60^\circ} = 0.076(1)$ from this subtraction method involving subsystems with zigzag edges is indeed seen to be consistent with the 5-component prediction $b_{60^\circ} = 0.074$.
We recall that in case of triangular bipartitions with zigzag edges, the  extracted value $b_{60^\circ}$ by direct fit had been found to be greater than the CFT value (Sec.~\ref{subsec:corners}).

Encouraged by this improvement, we applied a similar subtraction scheme to extract the $120^\circ$ corner coefficient.
In Fig.~\ref{fig:subtraction}(b), we show a corresponding pair of subsystems on a $6 \times 6$ lattice.
The light green subsystem contains four $120^\circ$ and two $60^\circ$ corners, while the combined subsystem has two $60^\circ$ and two $120^\circ$ corners.
Again, all edges in both subsystems are of the zigzag type. 
Since both subsystems have the same perimeter and the same number of $60^\circ$ corners, the edge related contributions and the logarithmic terms from the $60^\circ$ corners cancel. 
We therefore now expect the dominant scaling
\begin{equation}
\Delta S_{\text{corner}} \simeq 2 b_{120^\circ} \ln L ,
\label{eq:deltaS_120}
\end{equation}
However, as shown in Fig.~\ref{fig:subtraction}(d), the corresponding $\Delta S_{\text{corner}}$ does not exhibit a clear linear dependence on $\log L$. 
Thus we are unable to extract $b_{120^\circ}$ reliably through this method.
The reason for the failure of the subtraction method in this case is a mystery. 
Interestingly, even the success of the method in the $60^\circ$ case previously is also somewhat unexpected from the perspective discussed in Sec.~\ref{sec:more_discussion} of the main text.


\end{document}